\documentclass[fleqn,usenatbib]{mnras}

\usepackage{comment} 

\usepackage[T1]{fontenc}
\usepackage{ae,aecompl}

\usepackage{graphicx}	% Including figure files
\usepackage{amsmath}	% Advanced maths commands
\usepackage{amssymb}	% Extra maths symbols

\usepackage{xcolor}
\usepackage{ulem}

\usepackage{makecell}
\usepackage{multirow}
\usepackage{soul}
\usepackage{subcaption}

\usepackage{newtxtext,newtxmath}
\newcommand{\Ayr}{A_{\rm{yr}^{-1}}}

\title[GW stochastic background estimates from MB-II]{An estimate of the stochastic gravitational wave background from the MassiveBlackII simulation}

\author[ ]{Bailey Sykes,$^{1,2}$\thanks{E-mail: bailey.sykes@monash.edu}
Hannah Middleton,$^{1,3,4}$
Andrew Melatos,$^{1,4}$
Tiziana Di Matteo,$^{4,5}$ \newauthor
Colin DeGraf,$^{6}$
Aklant Bhowmick,$^{7}$
\\
$^{1}$School of Physics, University of Melbourne, Parkville, VIC, 3010, Australia\\
$^{2}$School of Physics and Astronomy, Monash University, Clayton, VIC, 3800, Australia \\
$^{3}$Centre for Astrophysics and Supercomputing, Swinburne University of Technology, Hawthorn, Victoria 3122, Australia \\
$^{4}$OzGrav-Melbourne, Australian Research Council Centre of Excellence for Gravitational Wave Discovery, Australia\\
$^{5}$McWilliams Center for Cosmology, Department of Physics, Carnegie Mellon University, Pittsburgh, PA 15213, USA \\
$^{6}$Institute of Astronomy and Kavli Institute for Cosmology, University of Cambridge, Madingley Road, Cambridge CB3 0HA, UK \\
$^{7}$Dept. of Physics, University of Florida, Gainesville, FL 32611, USA\\
}

\date{Accepted XXX. Received YYY; in original form ZZZ}

\pubyear{2021}

\begin{document}
\label{firstpage}
\pagerange{\pageref{firstpage}--\pageref{lastpage}}
\maketitle

\begin{abstract}
A population of super-massive black hole binaries is expected to generate a stochastic gravitational wave background (SGWB) in the pulsar timing array (PTA) frequency range of $10^{-9}$--$10^{-7}\,{\rm Hz}$. Detection of this signal is a current observational goal and so predictions of its characteristics are of significant interest. In this work we use super-massive black hole binary mergers from the MassiveBlackII simulation to estimate the characteristic strain of the stochastic background. We examine both a gravitational wave driven model of binary evolution and a model which also includes the effects of stellar scattering and a circumbinary gas disk. Results are consistent with PTA upper limits and similar to estimates in the literature. The characteristic strain at a reference frequency of $1\,{\rm yr}^{-1}$ is found to be $\Ayr = 6.9 \times 10^{-16}$ and $\Ayr = 6.4 \times 10^{-16}$ in the gravitational-wave driven and stellar scattering/gas disk cases, respectively. Using the latter approach, our models show that the SGWB is mildly suppressed compared to the purely gravitational wave driven model as frequency decreases inside the PTA frequency band.
\end{abstract}

\begin{keywords}
gravitational waves -- black hole mergers -- quasars: supermassive black holes -- galaxy: kinematics and dynamics
\end{keywords}

%%%%%%%%%%%%%%%%%%%%%%%%%%%%%%%%%%%%%%%%%%%%%%%%%%

%%%%%%%%%%%%%%%%% BODY OF PAPER %%%%%%%%%%%%%%%%%%

\section{Introduction}
Current wisdom holds that most galaxies host a super-massive black hole (SMBH)~\citep{KormendyHo:2013} whose mass ranges from $10^5\,{\rm M_{\odot}}$ to $10^{10}\,{\rm M_{\odot}}$. Hierarchical formation scenarios imply that galaxy mergers are frequent and the Universe contains a population of SMBH binaries (SMBHBs)~\citep{WhiteRees:1978,Begelman:1980}. These systems are predicted to emit gravitational waves (GWs) whose frequencies range from $10^{-9} \,{\rm Hz}$ to $10^{-7} \,{\rm Hz}$ and are accessible to modern pulsar timing arrays (PTAs)~\citep{FosterBacker:1990}. The incoherent superposition of these signals forms a stochastic GW background (SGWB), which is yet to be detected.

Predictions of the power spectrum of the SGWB have been made based on models of SMBHB populations. They suggest that the detection of the SGWB is within reach of modern PTAs. Theoretical work by~\citet{Phinney:2001} finds that the characteristic strain of a population of GW-driven, circular SMBHBs is a power law in frequency $h_{\rm c}(f) = \Ayr (f / 1 {\rm yr}^{-1})^{-2/3}$, where $\Ayr$ is the strain at frequency $f={\rm yr}^{-1}$.
The value of $\Ayr$ is determined by the distribution of the SMBHB population in redshift $z$ and chirp mass $\mathcal{M} = (m_{1}m_{2})^{3/5} / (m_{1} + m_{2})^{1/5}$
(where $m_1$, $m_2$ are the binary component masses, with $m_{1} > m_{2}$). 

PTAs are the tool of choice when searching for the SGWB. 
Current PTAs are the Parkes PTA~\citep[PPTA,][]{ShannonEtAlPPTA:2015}, the European PTA~\citep[EPTA,][]{DesvignesEtAlEPTA:2016}, the North American Nanohertz Observatory for Gravitational Waves~\citep[NANOGrav,][]{AlamEtAlNANOGrav:2020}, and the Indian PTA~\citep[InPTA,][]{JoshiInPTA:2018}.
In an effort to combined data and improve sensitivity, the individual PTAs work together as the International PTA~\citep[IPTA,][]{PereraEtAlIPTA:2019}. 
No conclusive detection has been made to date, however recent work by the NANOGrav group suggests a speculative SGWB signal with $\Ayr = 1.37-2.67 \times 10^{-15}$ \citep{Nanograv_sep:2020}. A further signal with $\Ayr = 1.9 - 2.6 \times 10^{-15}$ is reported by the PPTA \citep{PPTA:2021}. The EPTA too have announced the detection of a signal with $\Ayr = 2.23 - 3.84 \times 10^{-15}$ \citep{EPTA_2021}, likewise for the IPTA with $\Ayr = 2.0 - 4.0 \times 10^{-15}$ \citep{IPTA_2022}; however, none of these meet the criteria for a confirmed detection of the SGWB. Previously, upper limits have been reaching astrophysically interesting sensitivities and several studies have been made on their astrophysical interpretation~\citep[including][]{Chen:2019,ChenEtAlInterp:2017,NANOGrav:2018,MiddletonEtAl:2018,Middleton:2016,ShannonEtAlPPTA:2015}.

Existing estimates of the SGWB are based primarily on synthesised SMBHB populations produced by approaches broadly categorised as either cosmological hydrodynamical simulation or semi-analytic modelling. Cosmological hydrodynamical simulations directly solve for the dynamics of dark matter and gas, coupled with sub-resolution physics models to capture additional baryonic processes such as star formation, black hole growth and feedback \citep{SchayeCrain:2015, Illustris:2015, MBIIKhandaiEtAl:2015}. In contrast, semi-analytic models do not solve for the hydrodynamics of gas; instead, they generally rely on using analytic prescriptions to capture the baryonic processes within halos along their merger trees derived from N-body simulations \citep{Millenium:2005} or using the Press-Schechter formalism \citep{WyitheLoeb:2003, McWilliamsEtAl:2014}. The Illustris hydrodynamical simulations \citep{Illustris:2015} have been used to estimate $\Ayr = 0.6 \times 10^{-15}$ \citep{Kelley:2016}. This is done by taking simulated mergers and modelling the binary hardening effects for each one to make a prediction of their GW emission. Semi-analytic modelling of galaxy populations have produced similar estimates of $\Ayr < 4.6 \times 10^{-15}$~\citep{WyitheLoeb:2003},  $\Ayr = 4.1 \times 10^{-15}$~\citep{McWilliamsEtAl:2014} and $\Ayr = 3.5 - 15 \times 10^{-16}$ \citep{Sesana:2013b}. Estimates of the SGWB based on either method produce results of a comparable magnitude, however semi-analytic methods tend to be slightly higher. Reliable estimates of the SGWB guide PTA projects in regards to their target sensitivity \citep{Taylor:2016} and provide valuable information for generating initial constraints on astrophysical populations which relate to GW sources.

In this paper, a new estimate of $\Ayr$ is generated from the recent MassiveBlack-II simulation~\citep[MB-II,][]{MBIIKhandaiEtAl:2015}. The goal is to compare with previous estimates and thereby update the prospects of a detection. Estimates based on simulations are subjected to inherent uncertainties associated with their implementation of various physics models. For instance, \citet{Sesana:2009} uses merger trees from the Millenium simulation \citep{Millenium:2005} to synthesise a SMBHB population. In this case, the effects of SMBHs on galaxy evolution are not directly simulated, and are instead approximated by semi-analytic models. \citet{Kelley:2016} produce an estimate from the Illustris simulation \citep{Illustris:2015} which is comparable to MB-II in both number of particles and volume (MB-II is larger by only $58\%$) but suffers from the same limitations. Furthermore, the prediction by \citet{KulierCen:2015} is based on Enzo \citep{Enzo:2014} which has a $73\%$ larger simulation volume than MB-II however also has a very low resolution in most of that region, with just two small high-resolution volumes of order $10^{4}\,{\rm Mpc^{3}}$ embedded within it. MB-II, Illustris and Enzo have many features in common, including the sub-resolution physics which they attempt to model, although there are some exceptions, e.g. Enzo lacks active galactic nuclei (AGN) feedback. The method with which each simulation models the relevant physics differs and hence the comparison of the three (and others) provides insight into these phenomena.
At its resolution, MB-II is amongst the largest volume simulations to reach $z\approx0$; it reproduces a variety of observational constraints for the galaxy and black hole populations \citep{MBIIKhandaiEtAl:2015, Bhowmick:2019}. Additionally, MB-II implements a unique set of sub-resolution physics models compared to those used in previous work. These make MB-II an excellent candidate for making SGWB estimates, which will be a valuable addition to the existing literature.

This paper is structured as follows: in Section~\ref{sec:data} we outline the MB-II simulation and specify the data which is used in the analysis. Section~\ref{sec:method} details our approach to calculating SGWB predictions. In Section~\ref{sec:results} we present estimates of the characteristic strain of the SGWB from MB-II and place them in the context of the current estimates and observational upper limits. We discuss limitations of the methods and offer some concluding remarks in Section~\ref{sec:conclusion}.

\section{Data}
\label{sec:data}
In this work we use data from MB-II~\citep{MBIIKhandaiEtAl:2015}, an N-body hydrodynamical simulation offering a large simulation volume and high spacial resolution in a Lambda Cold Dark Matter ($\Lambda$CDM) cosmology. Like many modern cosmological simulations, it is based on a variation of GADGET-2~\citep{Springel:2005}. MB-II simulates $2\times 1792^{3}$ particles from redshift $z=159$ to $z=0$ in a cubic comoving volume $V_{\text{sim}}=(100\,{\rm Mpc/h})^{3}$ where $h=0.701$ is the dimensionless Hubble parameter. It simulates both the structure of the universe on very large scales and the formation of objects on comparatively smaller scales. Improved algorithms, in conjunction with modern high-performance computing facilities enable some of the largest volume and highest resolution simulation results to date; see \citet{MBIIKhandaiEtAl:2015} and references therein for more complete details of the implementation. Of particular interest is its ability to approximate SMBH formation, accretion and mergers, as this allows characterisation of SMBHB distributions in terms of both redshift and chirp mass.

The physical processes simulated in MB-II that are relevant to our analysis of the SGWB are summarised here. A SMBH of mass $5 \times 10^{5}\, {\rm h^{-1} M_{\odot}}$ is seeded in dark matter halos of mass at least $5 \times 10^{10}\, {\rm h^{-1}M_{\odot}}$ that form from density perturbations in the simulation; these objects have an accretion rate given by,
\begin{equation}
    \dot{M}_{\rm BH} = \frac{4 \pi G^{2} M^{2}_{\rm BH} \rho_{g}}{(c_{s}^{2} + v_{\rm BH}^{2})^{3/2}} \,
\end{equation}
where $\rho_{g}$ is the density and $c_{s}$ the sound speed of interstellar gas, $v_{\rm BH}$ is the peculiar velocity of a SMBH and $G$ is Newton's constant. The accretion rate is limited to twice the Eddington rate to avoid artificially high accretion. Mergers between SMBHs are triggered when two SMBHs are within each other's smoothed particle hydrodynamics (SPH) kernel -- a distance corresponding to the SPH smoothing length, which is normally a few kpc -- and are travelling with a sufficiently small relative velocity (less than the local gas sound speed) to be captured by each other's gravity. SMBH inspiral is not directly simulated due to the dynamics occurring below the resolution of the simulation. Additionally, since MB-II makes a record of a merger occurring once the above merging conditions are satisfied (not at the time at which physical coalescence occurs) the actual duration of the binary's inspiral is not known. It is possible that hardening mechanisms are insufficient to cause the binary to merge within the Hubble time \citep{Sesana:2013a}, nor reach a separation $<0.01\,{\rm pc}$ where gravitational emissions dominate \citep{Armitage:2002}. In this work we make two estimates: the first assumes that all binaries coalesce, while the second includes environmental effects and a time delay between binary formation and coalescence which can lead to some binaries never merging.

Data pertaining to SMBHs from the MB-II simulation are in the form of a $25\,{\rm GB}$ Structured Query Language (SQL) database accessible through the SQL database engine SQLite. It contains records for $133\,836$ distinct SMBHs and $50\,671$ mergers \citep{Lopez:2011}. Of these, $43\,211$ mergers have sufficient data available to perform the required calculations. Three SMBHs begin to develop unrealistically large masses due to a numerical bug which approach $10^{90}\,{\rm M_{\odot}}$ in the worst case at redshifts less than $z=0.6$. Owing to this being unphysical, any record or interaction containing the unique black hole identification (ID) number of these three are excluded from subsequent analyses. We use binary merger data to develop models of SMBHB populations. These data are obtained by querying all merger events and extracting the following at the time recorded for each merger: cosmic scale factor, primary and secondary masses, the accretion rate onto each SMBH and their unique ID numbers. These IDs are used solely to exclude misbehaving SMBHs while the scale factor, converted to redshift, and the chirp mass, calculated from constituent masses, are used to develop the population models. The velocity dispersion and stellar mass of host galaxies are obtained from a separate database and are used in the modelling of binary evolution.

Throughout this work, any cosmological parameters are chosen to be consistent with WMAP7 \citep{WMAP7}, the same cosmology used for MB-II.

\section{Calculating \texorpdfstring{$\MakeLowercase{\mathbf{h}_{\rm c}(f)}$}{hc(f)}}
\label{sec:method}
Here we describe our methodology for predicting $h_{\rm c}(f)$. Section~\ref{subsec:summation} demonstrates how $h_{\rm c}(f)$ is obtained by summing the contribution of each merger in MB-II assuming circular binaries with no environmental effects. In Section~\ref{subsec:hardening} we extend the method to include additional physics which may affect the SGWB spectrum: specifically the scattering of stars in the binary loss cone, torques with a circumbinary gas disk and a time delay between binary formation in MB-II and physical coalescence (Section~\ref{subsec:delay}).

\subsection{Merger sum}
\label{subsec:summation}
We treat each SMBHB merger in MB-II as a source of GWs and add them to find the total expected signal. Following \citet{Chen:2017}, and the earlier work of \citet{Phinney:2001}, we can write the dimensionless quantity $h_{\rm c}(f)$ expected from a population of SMBHBs as
\begin{equation} 
    \label{eqn:strainsq}
    h_{\rm c}^{2}(f) = \frac{4G}{\pi c^{2} f} \int_{0}^{\infty}dz \int_{0}^{\infty} d\mathcal{M} \frac{d^{2}n}{dz d\mathcal{M}} \frac{dE}{df_{\rm r}}\,,
\end{equation}
where $d^{2}n/dzd\mathcal{M}$ is the comoving differential number density of mergers per redshift $z$ interval $(z, z + dz)$ and chirp mass interval $(\mathcal{M}, \mathcal{M} + d\mathcal{M})$ in units of ${\rm Mpc}^{-3}$, $c$ is the speed of light, $f$ and $f_{\rm r}$ are the observed and the source rest frame GW frequency, respectively. The quantity $dE/df_{\rm r}$ denotes the energy emitted in the form of GWs per rest frame frequency interval. Furthermore, from~\citet{Phinney:2001}, for a circular SMBHB we have
\begin{equation}
    \label{eqn:energyfreq}
    \frac{dE}{df_{r}} = \frac{\pi^{2/3} G^{2/3} \mathcal{M}^{5/3}}{3 f_{\rm r}^{1/3}}\,.
\end{equation}
By transforming from the source rest frame to the observers frame using $f_{\rm r} = f(1+z)$ we obtain
\begin{equation}
    \label{eqn:strainsq_full}
    h_{\rm c}^{2}(f) = \frac{4G^{5/3}}{3 \pi^{1/3} c^{2} f^{4/3}} \int_{0}^{\infty}dz \int_{0}^{\infty} d\mathcal{M} \frac{d^{2}n}{dzd\mathcal{M}} \frac{\mathcal{M}^{5/3}}{(1+z)^{1/3}}\,.
\end{equation}
Again, following~\cite{Chen:2017}, we write the number density as a sum of delta functions, selecting particular pairs of $z$ and $\mathcal{M}$ associated with each SMBHB,
\begin{equation}
    \label{eqn:sum_num_density}
    \frac{d^{2}n}{dzd\mathcal{M}} = \frac{1}{V_{{\rm sim}}} \sum_{i=1}^{N}\delta(z-z_{i}) \delta(\mathcal{M} - \mathcal{M}_{i})\,,
\end{equation}
where $\delta$ is the Dirac delta function and $N$ is the number of mergers in the simulation data. Equation~\eqref{eqn:sum_num_density} is normalised to give $\int~dz\int~d\mathcal{M}\,d^{2}n/dzd\mathcal{M}=N/V_{{\rm sim}}$, where ${V_{\rm sim} = (100 \, {\rm Mpc/h)^{3}}}$ is the MB-II simulation volume. 

Since the simulation records mergers as occurring when the separation and relative velocity conditions outlined in Section \ref{sec:data} are satisfied, we do not know how the binary behaves when it is emitting GWs at lower binary separations; it could coalesce very quickly, or not at all. As an initial approximation, we assume here that all SMBHBs coalesce while maintaining a constant $\mathcal{M}$ and $z$, i.e. that they coalesce instantly once the binary forms. Upon substituting equation~\eqref{eqn:sum_num_density} into equation~\eqref{eqn:strainsq_full}, we find that the characteristic strain from a population of discrete sources, under the above approximations, can be written as
\begin{equation}
    \label{eqn:summation_hc}
    h_{\rm c}^{2}(f) = \frac{4G^{5/3}}{3 V_{{\rm sim}} \pi^{1/3} c^{2} f^{4/3}}  \sum_{i=1}^{N} \frac{\mathcal{M}_{i}^{5/3}}{(1+z_{i})^{1/3}}  \,.
\end{equation}

\subsection{Sum with hardening mechanisms}
\label{subsec:hardening}
We extend the method of the previous section to include the effects of the SMBHB scattering stars in the binary loss cone and momentum transfer to a circumbinary gas disk using the approach of \citet{Kelley:2016} to modify equation~\eqref{eqn:summation_hc},
\begin{equation}
    \label{eqn:summation_hardening}
    h_{\rm c}^{2}(f) = \frac{4G^{5/3}}{3 V_{{\rm sim}} \pi^{1/3} c^{2} f^{4/3}}  \sum_{i=1}^{N} \frac{\mathcal{M}_{i}^{5/3}}{(1+z_{i})^{1/3}} \frac{\tau_{{\rm h},i}(f)}{\tau_{{\rm gw}, i}(f)}  \,.
\end{equation}
We define the binary hardening time $\tau_{\rm h}(f) = a / (da/dt_{\rm r})$ where $a$ is the semimajor axis of the binary and $t_{\rm r}$ is the rest frame time. $\tau_{\rm h}$ refers to the sum of all mechanisms which drive the binary to merge while $\tau_{\rm gw}$ is the hardening time assuming a merger driven purely by gravitational radiation. In equation~\eqref{eqn:summation_hc} we implicitly assume that $\tau_{\rm h} = \tau_{\rm gw}$. Including these additional physical effects to the model reduces  $h_{\rm c}(f)$ as $\tau_{\rm h} < \tau_{\rm gw}$. From this point onward, we differentiate the formation of the SMBHB from the coalescence of the SMBHs; the properties of the binary at its formation are given in the MB-II data while the properties at the moment of coalescence are determined by evolving the system in time.

Two effects are considered in addition to GW emission to drive SMBHBs to merge. Firstly, stars which approach the SMBHB undergo a three body interaction which ejects the star, removing energy and angular momentum from the binary \citep{Dotti:2012}. The fraction of stars in a galaxy which may undergo this scattering is relatively small and occupy a region known as the loss cone. This region may become depleted as stars are scattered with the binary which reduces the efficiency with which this mechanism is able to harden the binary. The loss cone is repopulated by diffusion of surrounding stars back into the loss cone where they are able to efficiently scatter \citep{Milosavljevic:2003, Makino:2004}. We consider the contribution to $da/dt_{\rm r}$ from \citet{Sesana:2013a},
\begin{equation}
    \label{eqn:scattering}
    \bigg( \frac{da}{dt_{\rm r}} \bigg)_{\rm scatter} = -\frac{a^{2} G \rho H}{\sigma}\,,
\end{equation}
where $\rho$ and $\sigma$ are the density and velocity dispersion of stars respectively, and H is a dimensionless hardening rate of approximately $15-20$ \citep{Dotti:2012}. 

We also consider the effect of a circumbinary disk on the binary, a process in which an accumulation of gas in the orbital plane of the binary extracts angular momentum from it \citep{Kocsis:2011, Ivanov:1999, Dotti:2012}. This gas is expected in the binary environment as SMBHBs form in galaxy mergers, which are known to cause inflows of gas towards the galactic nucleus \citep{Kocsis:2011}. The eccentricity of the binary is expected to be increased by these gas interactions; we do not consider this effect in our analysis. An equation governing the binary separation in a gas disk environment is given by \citet{Sesana:2013a},
\begin{equation}
    \label{eqn:circumbinary_disk}
    \bigg( \frac{da}{dt_{\rm r}} \bigg)_{\rm disk} = -\frac{2\dot{M} (m_{1} + m_{2})}{m_{1} m_{2}} \sqrt{a a_{0}}\,,
\end{equation}
where $\dot{M}$ is the accretion rate onto the disk and $a_{0}$ is the initial semimajor axis length at which the the mass of the local disk equals the mass of the secondary black hole.

The total rate of change of the semimajor axis $a$ is given by the sum of the effects from GW emission, stellar scattering, and the circumbinary disk, i.e.:
\begin{eqnarray}
    \label{eqn:total_da_dt}
    \bigg( \frac{da}{dt_{\rm r}} \bigg)_{\rm all} &=& -\frac{64G^{3} m_{1} m_{2} (m_{1} + m_{2})}{5 c^{5} a^{3}} \nonumber \\
    & & - \frac{a^{2} G \rho H}{\sigma} - \frac{2 \dot{M} (m_{1} + m_{2})}{m_{1} m_{2}} \sqrt{a_{0} a}\,.
\end{eqnarray}
The ratio $\tau_{\rm h} / \tau_{\rm gw}$ can be written simply as, 
\begin{equation}
    \label{eqn:tau_frac}
    \frac{\tau_{{\rm h}}}{\tau_{{\rm gw}}} = \frac{\bigg(\frac{da}{dt_{r}} \bigg)_{\rm gw}}{\bigg(\frac{da}{dt_{r}} \bigg)_{\rm all}}
\end{equation}
By noting that the GW frequency from the binary is twice the orbital frequency \citep{Hilborn:2018} and applying Kepler's third law, we can arrive at an equation for the semi-major axis $a$,
\begin{equation}
    \label{eqn:kepler}
    a = \frac{G^{1/3}(m_{1} + m_{2})^{1/3}}{(\pi f_{\rm r})^{2/3}}\,
\end{equation}
which allows us to write equation~\eqref{eqn:tau_frac} as a function of frequency. This allows it to be used in equation~\eqref{eqn:summation_hardening} by transforming to the observer's frame with $f_{\rm r} = (1+z)f$.

For each merger, the values of $\rho$, $\sigma$, $\dot{M}$ and $a_{0}$ must be determined. We take $H=15$ exactly for these calculations~\citep{Dotti:2012}. The velocity dispersion, $\sigma$, is obtained from MB-II data at the next snapshot following the binary formation. This inherently assumes that the velocity dispersion does not change significantly before the snapshot. As a matter of fact, once the binary separation reaches a few parsecs, at which point it becomes important in equation \eqref{eqn:scattering} for the scattering mechanism, the velocity dispersion has stabilised \citep{Stickley:2013}. This renders the assumption valid under the condition that there are no interactions with a third galaxy before the snapshot.

Further data from MB-II are extracted to build interpolants of the accretion rate, $\dot{M}(z)$, of each SMBH throughout the simulation. These permit an approximation of the accretion rate onto the binaries at any time throughout their evolution, including the period after binary formation but before coalescence.

The hardening effects described above are relevant on scales below the resolution of MB-II and as such, simulation data may be too coarse to provide an accurate estimate of certain parameters. To approximate the density $\rho$, we follow \citet{Chen:2017} who use the density profile of \citet{Dehnen:1993} and the $a_{\rm c}-M_{*}$ \citep{Dabringhausen:2008} and $M_{\rm BH}-M_{*}$ \citep{KormendyHo:2013} relations, where $M_{*}$ is galaxy bulge mass and $a_{\rm c}$ is its characteristic radius, to find the galaxy density at the binary influence radius,
\begin{equation}
    \label{eqn:rho_density}
    \rho = 0.092\,{\rm M_{\odot}\,pc^{-3}} \mathscr{F}(\gamma) \bigg( \frac{m_{1} + m_{2}}{10^{9}\,{\rm M_{\odot}}} \bigg)^{\mathscr{G}(\gamma)} \,, 
\end{equation}
where
\begin{equation*}
    \mathscr{F}(\gamma) = \frac{(3 - \gamma) 92^{\gamma / (3 - \gamma)}}{(2^{1 / (3 - \gamma)} - 1)^{3}}\,,
\end{equation*}
\begin{equation*}
    \mathscr{G}(\gamma) = -0.68 - 0.138 \frac{\gamma}{3 - \gamma}\,,
\end{equation*}
and $0.5 < \gamma < 2$ \citep{Chen:2017} is a dimensionless parameter of the galaxy density model.

We identify $a = a_{0}$ as the semi-major axis of the binary under the condition,
\begin{equation}
    \label{eqn:a0_eq}
    \frac{4 \pi r_{0}^{2} \Sigma_{0}}{\mu} = 1
\end{equation}
where we take $r_{0} = 1.5 a_{0}$ \citep{Kocsis:2011}, $\mu = m_{1}m_{2} / (m_{1} + m_{2})$ is the reduced mass, and $\Sigma_{0}$ is the disk surface density. We use \citet{Ivanov:1999} to approximate the surface density,
\begin{equation}
    \label{eqn:surface_density}
    \Sigma_{0} \approx 10^{6} ( 100 \alpha )^{-4/5} \bigg( \frac{m_{1} + m_{2}}{10^{8}\,{\rm M_{\odot}}} \bigg)^{1/5} \bigg(\frac{100 \dot{M}}{\dot{M}_{\rm Edd}} \bigg)^{3/5} a^{-3/5} \,{\rm kg\,m^{-2}},
\end{equation}
in which $\alpha$ is a parameter describing the disk shape, which is loosely constrained to $0.01-1$ by numerical studies \citep{Kocsis:2011}.

There is some freedom to chose values for $\gamma$ in equation~\eqref{eqn:rho_density} and $\alpha$ in equation~\eqref{eqn:surface_density} owing to the natural variation in galaxy and gas disk properties. The ranges specified for each parameter are chosen to be as broad as is physically reasonable. Uniform and log-uniform distributions are used to sample the parameter space for $\gamma$ ($0.5$--$2$) and $\alpha$ ($0.01$--$1$) respectively, with a random pair of values assigned to each SMBHB. We repeat the calculation $50$ times, regenerating random parameter sets for each SMBHB each time to explore the variation in the SGWB estimate that this randomness produces.

\subsection{Coalescence delay and mass gain}
\label{subsec:delay}
Recall that MB-II does not record when a SMBHB actually coalesces, but records a merger when two SMBHs are within a few kiloparsecs. As such, there should be a delay between the recorded merger and the emission of GWs near coalescence. It is also possible that SMBHBs evolve too slowly to coalesce within the Hubble time, meaning that some or all the GW contribution of a given SMBHB may be absent. Furthermore, provided that SMBHBs experience a time delay as the system hardens, the component masses will also change owing to accretion during that period. These two effects are included in the calculation described in Section~\ref{subsec:hardening}. The approach is outlined here.

The delay before coalescence is modelled in two parts: the first part consists of time under the influence of dynamical friction and the second part for time spent evolving under other mechanisms. This separation is made because, at the initial binary separation of a few kpc, the dominant mechanism driving the evolution of the binary is dynamical friction. In this regime, the GW emissions of the binary are minimal, and well outside the $10^{-9}\, \rm Hz - 10^{-7}\, \rm Hz$ frequency range we are interested in. As such we elect to simplify the calculation and simply shift the binary forward in time by some amount \citep{Ryu:2018},
\begin{equation}
    \label{eqn:t_df}
    t_{\rm df} \approx 3\, {\rm Gyr}\, \bigg( \frac{6}{\ln \Lambda} \bigg) \bigg( \frac{r}{5 \, {\rm kpc}} \bigg)^{2} \bigg(\frac{\sigma}{200 \, {\rm km \, s^{-1}}} \bigg) \bigg( \frac{m_{2}}{10^{8} \, M_{\odot}} \bigg)^{-1} \,.
\end{equation}
Here, $\ln \Lambda$ is the Coulomb logarithm, $r$ is the initial binary separation, $\sigma$ is the galactic velocity dispersion and $m_{2}$ is the mass of the satellite (lower mass) SMBH. Similar equations are found in \citet{Begelman:1980} and \citet{GalacticDynamicsBinney:2008}, for example, but this one is chosen as all the required quantities are known, or simple to calculate. We adopt an approximation of the Coulomb logarithm following \citet{Boylan:2007} and \citet{Dosopoulou:2017} where $\ln \Lambda = \ln{1 + M_{\rm host} / m_{\rm 2}}$, where $M_{\rm host}$ is the remnant galaxy stellar mass.

During the second part of the delay the hardening mechanisms detailed in Section~\ref{subsec:hardening} become significant and the binary is evolved in time from an initial separation of about a parsec. Leveraging the assumption of circular binaries, this corresponds to a GW frequency of $10^{-10}\,{\rm Hz}$: below this frequency, we do not calculate the SWGB. Starting at this minimum frequency, the time elapsed (in the source frame) as the binary separation decreases, and the GW frequency increases is calculated as,
\begin{equation}
    \label{eqn:dtdf}
    \Delta t_{\rm r} = \frac{dt_{\rm r}}{da} \frac{da}{df} \Delta f \,,
\end{equation}
which is readily calculable from equations \eqref{eqn:total_da_dt} and \eqref{eqn:kepler}. The resolution of the numerical integration is governed by $\Delta f$, the frequency step size. The total time delay is then $t_{\rm df} + \sum \Delta t_{\rm r}$ where the sum is over each frequency step from $10^{-10} \, {\rm Hz}$ to $10^{-7} \, {\rm Hz}$; we use $100$ steps over log-spaced intervals. If, at any time during its evolution, a binary evolves past the age of the Universe, the evolution is halted. This means that, if the dynamical friction driven infall takes too long, some systems may never emit GWs. Additionally, if stellar scattering, gas disk interactions or GW emission are too slow to induce coalescence, the simulation may end while the system is emitting in the $10^{-10}$--$10^{-7} \, {\rm Hz}$ range; this can result in the SWGB contribution of that system being truncated above some frequency.

The component masses of the binary increase due to accretion prior to coalescence. This is expected to have a non-negligible impact on the SMBHB evolution and lifetime \citep{Siwek:2020}. To account for this, as the time delay is updated with each frequency step, the product of the interpolated accretion rate at the new time and $\Delta t$ is added to the current SMBH mass ($M_{\rm BH} + \dot{M}(t) \Delta t$). This occurs for both SMBHs in the binary.

\section{SGWB predictions}
\label{sec:results}
In this section we present our SGWB predictions and compare them to current observational constraints from the PTAs. 
The $95\%$ upper limits from the individual PTAs are $\Ayr < 1.45 \times 10^{-15}$, $\Ayr < 1.0 \times 10^{-15}$, and $\Ayr < 3.0 \times 10^{-15}$ for NANOGrav~\citep{NANOGrav:2018}, the PPTA~\citep{ShannonEtAlPPTA:2015}, and the EPTA~\citep{EPTASGWB:2015} respectively. A similar value of $\Ayr < 1.7 \times 10^{-15}$ is reported by the IPTA~\citep{IPTA:2016}.
The 2015 PPTA upper limit is the most stringent and it is used in this work to provide context to the results. 

\subsection{Estimates of \texorpdfstring{$\Ayr$}{Ayr}}
\begin{figure*}
    \centering
    \includegraphics[width = \textwidth]{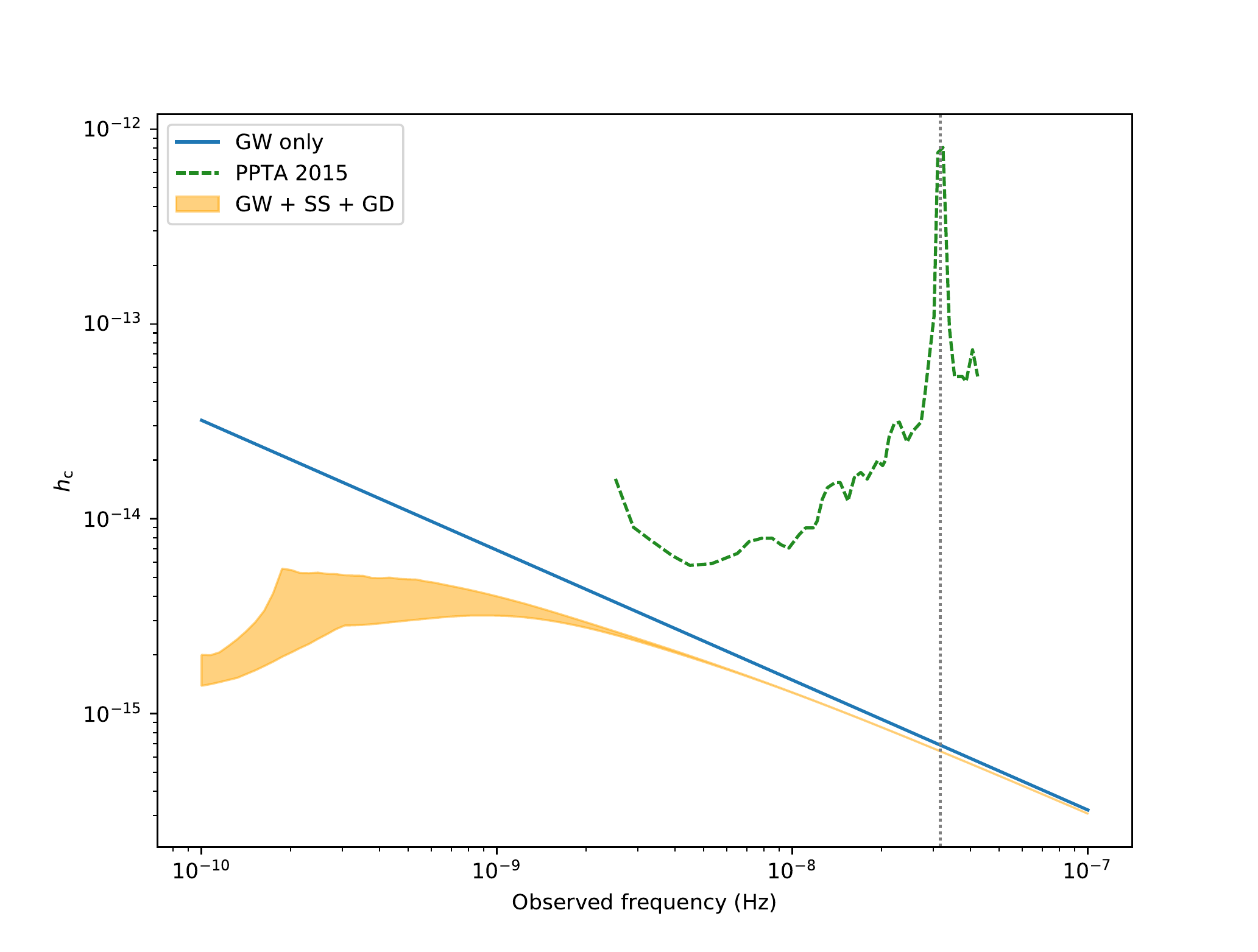}
    \caption{Characteristic strain for the SGWB from MB-II assuming a purely GW-driven evolution in blue and also including stellar scattering (SS) and the torques of a circumbinary gas disk (GD) in orange. The orange shaded region encompasses 50 realisations of the SGWB with hardening mechanisms and coalescence delay. These curves are compared to the 2015 PPTA~\citep{ShannonEtAlPPTA:2015} upper limit (green-dashed). The grey dotted vertical line shows the position of $f = 1\,{\rm  yr^{-1}}$.}
    \label{fig:strain}
\end{figure*}

Fig.~\ref{fig:strain} collates our main results in calculating the characteristic strain, $h_{\rm c}(f)$, expected from the population of SMBHBs in the MB-II data according to the methods outlined in Section \ref{sec:method}. The first method (Section \ref{subsec:summation}), which includes only gravitational hardening mechanisms, gives $h_{\rm c} \propto f^{-2/3}$ and is shown by the blue line in Fig.~\ref{fig:strain}. The method which includes stellar scattering,  gas interactions (Section~\ref{subsec:hardening}) and a time delay between binary formation and coalescence (Section~\ref{subsec:delay}) deviates from a pure power law by a factor of at least 2 at frequencies below $6.4 \times 10^{-10}\,{\rm Hz}$. It is shown by the yellow curve in Fig.~\ref{fig:strain}.

Both methods produce results which are consistent with the PPTA upper limit shown by the green-dashed line in Fig.~\ref{fig:strain}, i.e., they lie below the $95\%$ upper limit. The GW-only method produces $\Ayr= 6.9 \times 10^{-16}$ which differs by less than $1.5\%$ to the corresponding GW-only estimate of $A_{{\rm yr}^{-1}} = 7 \times 10^{-16}$ from the Illustris simulations \citep{Kelley:2016}.

Including the additional effects of stellar scattering and circumbinary disk interactions produces $\Ayr = 6.4 \times 10^{-16}$. The characteristic strain is significantly suppressed at lower frequencies compared to $h_{\rm c} \propto f^{-2/3}$. At higher frequencies, in the range of $10^{-8} \, {\rm Hz}$--$10^{-7} \, {\rm Hz}$, the signal is less suppressed, and approaches the GW-only estimate as frequency increases further. PTAs typically are sensitive to frequencies between $10^{-9}\,{\rm Hz}$ and $10^{-7}\, {\rm Hz}$ which means that a diminished signal due to stellar and gas interactions may make the SGWB more difficult for PTAs to detect~\citep{ShannonEtAlPPTA:2015}. We find a maximum suppression by a factor of $1.7$ at the low end of the PTA band: $10^{-9}\,{\rm Hz}$.
The turnover in the characteristic strain spectrum is at $\sim 6 \times 10^{-10}\,{\rm Hz}$. This is a comparable frequency to that predicted by others. \citet{Sesana:2013a} use the same models as this work, but also include eccentricity, and find turnovers between $7 \times 10^{-10}\,{\rm Hz}$ and $3 \times 10^{-9}\,{\rm Hz}$. \citet{Chen:2017} use a parameterised SMBHB population model and consider the effects of eccentricity and stellar scattering, but not gas disk interactions; they find a turnover frequency of $\sim 3 \times 10^{-10}\,{\rm Hz}$. \citet{Kelley:2016} consider the effects of dynamical friction, loss cone stellar scattering and a circumbinary gas disk, and find a turnover at $\sim 3 \times 10^{-10}\,{\rm Hz}$.

Gas disk interactions cause greater suppression of the SGWB signal at higher frequencies (by a factor of $\sim 10$) than suppression caused by stellar scattering. As such, we find that the turnover position, when the effects are combined, is dominated by the effects of gas disk interactions. The model we use for gas-driven binary evolution does not take into account the decrease in efficiency of the momentum transfer from the binary to the disk as the binary separation gets smaller \citep{Haiman:2009}. Ignoring this means that $da/dt_{\rm r}$ is overestimated, especially at lower separations where the GWs are emitted at a higher frequency. This may partially explain minor differences between our estimate of the SGWB and other predictions. Using a model such as in \citet{Kocsis:2011}, where the binary-disk evolution is modelled more precisely may produce results in closer agreement with other works.

The orange region in Fig.~\ref{fig:strain} is produced by sampling the $\alpha$ and $\gamma$ distributions for each SMBHB as per Section \ref{subsec:hardening}. The variation which this induces between individual estimates is shown by the width of the region at a given frequency. In general, the influence of variation in these parameters decreases as the GW frequency increases; at the low end of the PTA frequency band, $10^{-9} \, {\rm Hz}$, the width is $0.1 \, {\rm dex}$ increasing to a maximum of $0.45 \, {\rm dex}$ at $1.9 \times 10^{-10} \, {\rm Hz}$. The region is peaked or kinked along its boundary due to massive SMBHBs which dominates the signal. Consider the peak at $2 \times 10^{-10} \, {\rm Hz}$: this is the result of one massive SMBHB system with very strong GW emissions. The stronger suppression of $h_c$ below $2 \times 10^{-10} \, {\rm Hz}$ indicates that the binary hardened quickly at these lower frequencies/higher binary separations. The peak indicates a sudden slowing of the hardening rate of that SMBHB. Exploring the data in more detail shows that this corresponds to a sudden decrease in the accretion rate of a component SMBH. A lower accretion rate decreases the efficiency of gas-disk interactions, as per equation \eqref{eqn:circumbinary_disk}, resulting in more time spent emitting GWs at the frequency of the peak and hence a stronger signal. The accretion rate of that particular SMBH increases again quite smoothly after this event, decreasing the SGWB contribution accordingly at later times. Removing the SMBH associated with this peak produces a characteristic strain curve which is smoother, but also weaker. One should be careful that no SMBHs such as those which were removed for being unrealistically massive (see Section \ref{sec:data}) remain in the data; the SMBH discussed here may be a candidate for removal, however its growth pattern indicates that it is merely a statistically rarity.

The mergers present in the MB-II data represent a single realisation of infinitely many possible evolution pathways for the simulation volume. If the simulation is re-run with a different initial random seed, one can expect some variation in the population of SMBHBs and the resulting SGWB. Statistically rare events such as mergers with $\mathcal{M} > 10^{9} \, M_{\odot}$ may occur in higher or lower numbers, with that random variation in a few events producing a large change in the final prediction. We perform an approximate test of this variation by regenerating SMBHB component masses via an $M_{\rm BH} - M_{*}$ relation.

Observational studies typically estimate the intrinsic scatter of a given $M_{\rm BH} - M_{*}$ relation at around $0.3\,{\rm dex}$ \citep{KormendyHo:2013, McConnell_Ma:2013}; in contrast, \citet{DeGraf:2015} have shown that MB-II exhibits a tighter $M_{\rm BH} - M_{*}$ relation with intrinsic scatter as low as $0.18\, {\rm dex}$ when $z < 1$ (see \citet{DeGraf:2015} for a more detailed discussion). We investigate the uncertainty in our MB-II estimate by adding the scatter from these $M_{\rm BH} - M_{*}$ relations into the MB-II data. The procedure we describe here results in a set of hypothetical re-runs of MB-II with different SMBH populations and resulting SGWB estimates, from which we can estimate the uncertainty in $\Ayr$. To produce one of these new SMBH populations we start from the original MB-II data and then, for each component SMBH in a merger, we generate a corresponding host galaxy mass $M_{*,{\rm estimate}}$ following an $M_{\rm BH} - M_{*}$ relation with no scatter. We then use the same relation -- this time with scatter -- to draw estimates of $M_{\rm BH, estimate}$ from these generated $M_{*, {\rm estimate}}$, thus simulating scatter in the SMBH masses: $M_{\rm BH, \, actual} \rightarrow M_{\rm *, \, estimate} \rightarrow M_{\rm BH, \, estimate}$. This procedure is done using the $M_{\rm BH} - M_{*}$ relations of both \citet{KormendyHo:2013} and \citet{DeGraf:2015} to quantify the impact of larger or smaller scatter on the uncertainty in $\Ayr$.

\begin{figure}
    \centering
    \includegraphics[width = \linewidth]{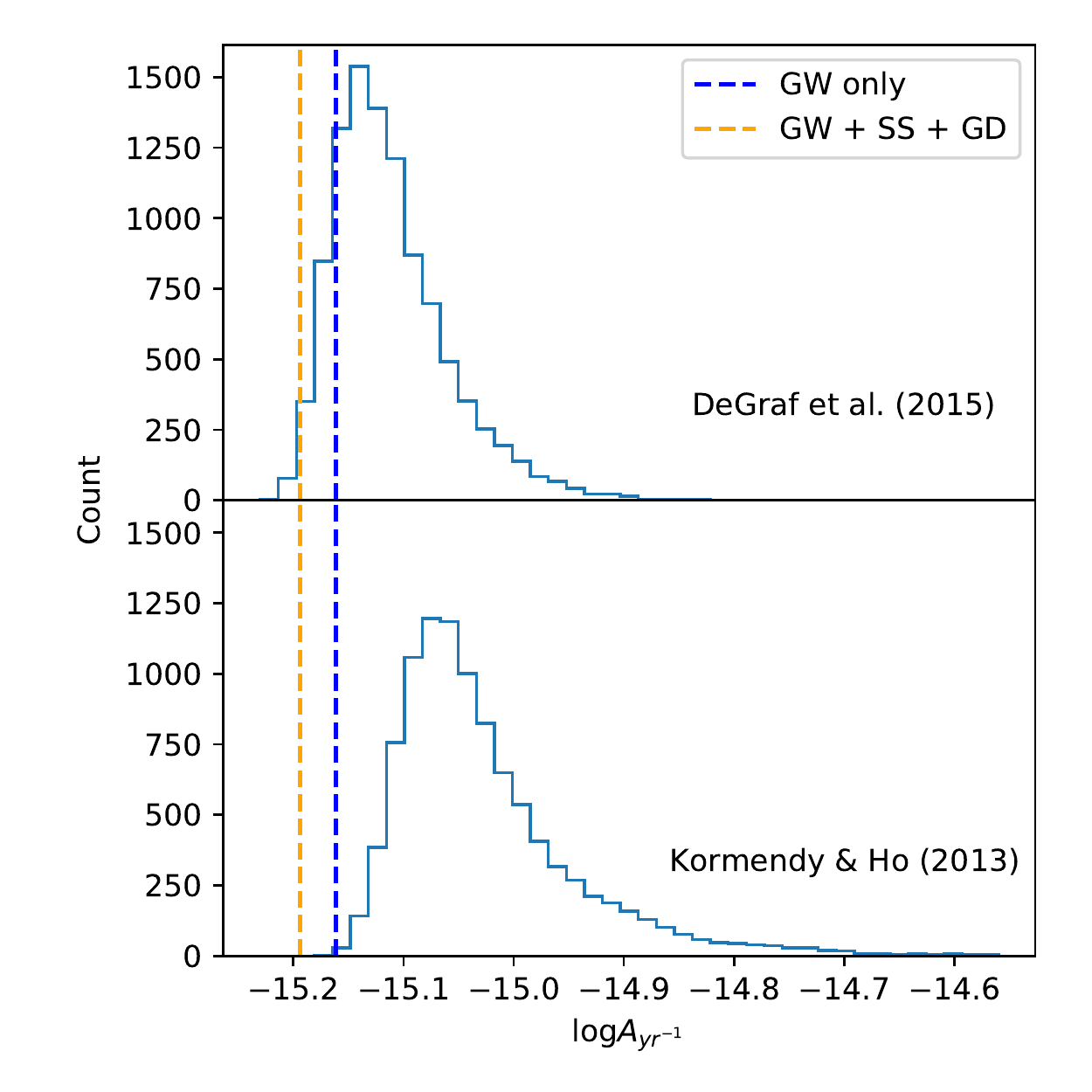}
    \caption{Distributions of $\Ayr$ estimated through the \citet{DeGraf:2015} (upper) and \citet{KormendyHo:2013} (lower) $M_{\rm BH} - M_{*}$ relations using the GW-only method described in the text. These distributions are used to assess the uncertainty of the MB-II $\Ayr$ estimates presented in this work. The horizontal axis is truncated on the right where there is less data. The values of $\Ayr$ estimated directly from MB-II data are shown as dashed vertical lines.}
    \label{fig:Ayr_dists}
\end{figure}

We generate $10\,000$ realisations of the SMBH population through the aforementioned method and use them in place of the original masses to recalculate $\Ayr$ in the GW-driven case. In effect, this corresponds to $10\,000$ re-runs of the simulation, or looking at $10\,000$ distinct but otherwise equivalent volumes. From SMBH masses generated through the $M_{\rm BH} - M_{*}$ relation of \citet{DeGraf:2015}, estimates of $\Ayr$ vary over a range of $0.5 \, {\rm dex}$. Similarly for the \citet{KormendyHo:2013} relation, the new estimates exhibit a range of $0.8 \, {\rm dex}$. The distributions of $\Ayr$ produced by these $10\,000$ estimates are shown in Figure \ref{fig:Ayr_dists}. Recall that the $10\,000$ estimates assume GW-driven binaries so it is not surprising that the MB-II GW-only estimate (blue dashed line) matches the distributions more closely than the estimate with environmental interactions (orange dashed line). We also note that the MB-II GW-only estimate lies at the lower edge of the~\cite{KormendyHo:2013} distribution (lower panel). This is because the~\cite{KormendyHo:2013} $M_{\rm BH} - M_{*}$ relation predicts a higher proportion of high mass SMBHs than many other $M_{\rm BH} - M_{*}$ relations and therefore also predicts a higher $\Ayr$ (as also seen in~\cite{MiddletonEtAl:2018}).
The ranges of the distributions are quite broad, however this is to be expected when, looking forward to Section \ref{sec:SGWBcontributions}, we find $50\%$ of the signal comes from just four SMBHBs. In a realisation of the simulation volume (or perhaps an adjacent volume) which does not include these four rare events or similar massive SMBHBs, the SGWB could be significantly diminished; it is also possible for a comparable volume to contain a greater number of such SMBHBs, resulting in a much larger SGWB.

The average lifetime of the MB-II SMBHBs -- the time between binary formation and coalescence -- is $3.3 \, {\rm Gyr}$. Individual SMBHBs have lifetimes between tens of Myrs up to $> 12 \, {\rm Gyr}$. The mean lifetime is consistent with some binary evolution models of \citet{Kelley:2016} and with \citet{Sesana:2015}, but is longer than in \citet{Chen:2019}. Some of this variation may be attributed to differences in the initial separations ($20$--$30 \, {\rm kpc}$ in \citet{Chen:2019} compared to our $1$--$2 \, {\rm kpc}$) however the primary cause is the difference between the binary evolution models, as reported in \citet{Siwek:2020}.

\subsection{Contributions to the SGWB}
\label{sec:SGWBcontributions}
\begin{figure}

\begin{subfigure}{0.5\textwidth}
\includegraphics[width=\linewidth]{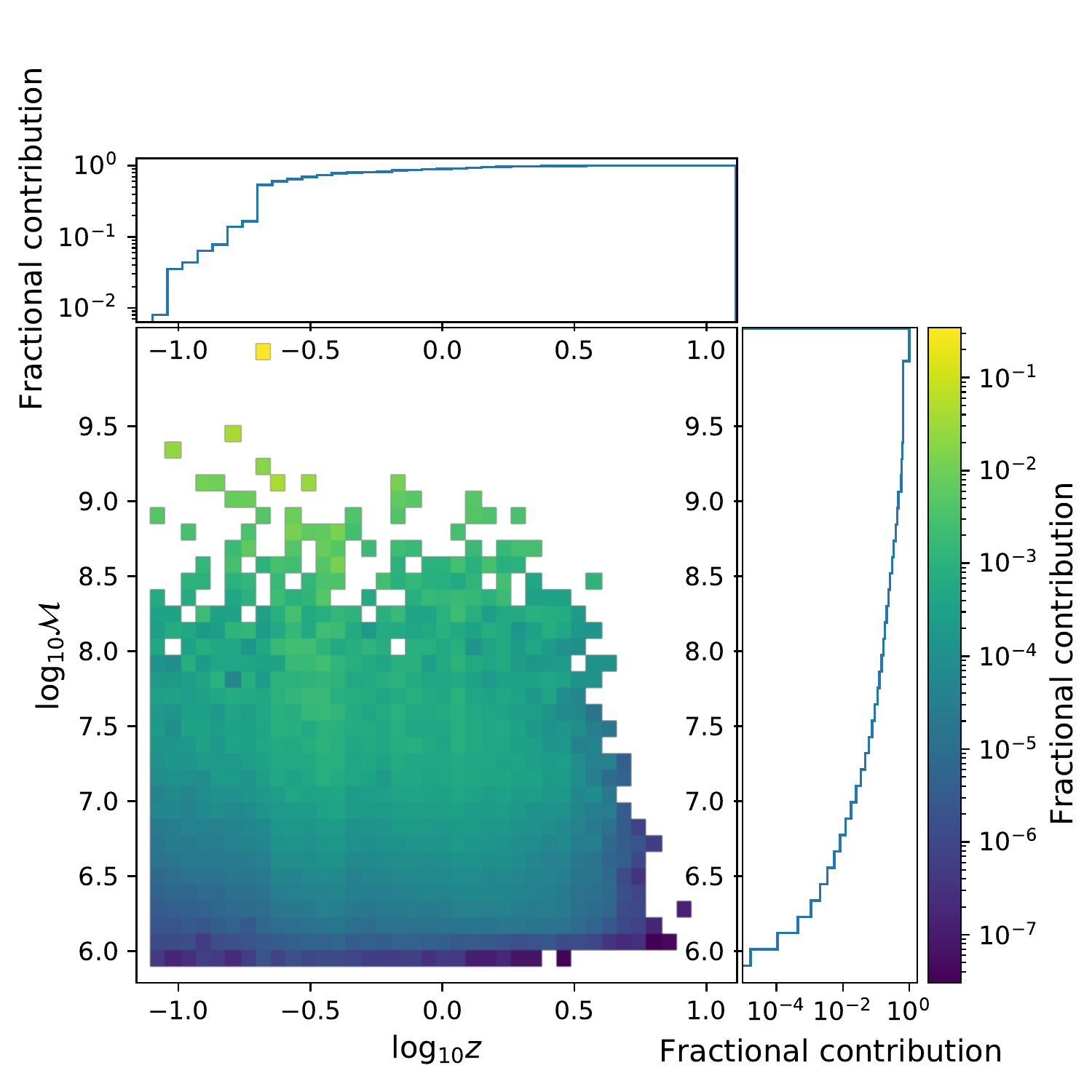} 
\caption{GW-driven at $f = 1\,{\rm yr}^{-1}$}
\label{fig:contrib_gw_1yr}
\end{subfigure}
\begin{subfigure}{0.5\textwidth}
\includegraphics[width=\linewidth]{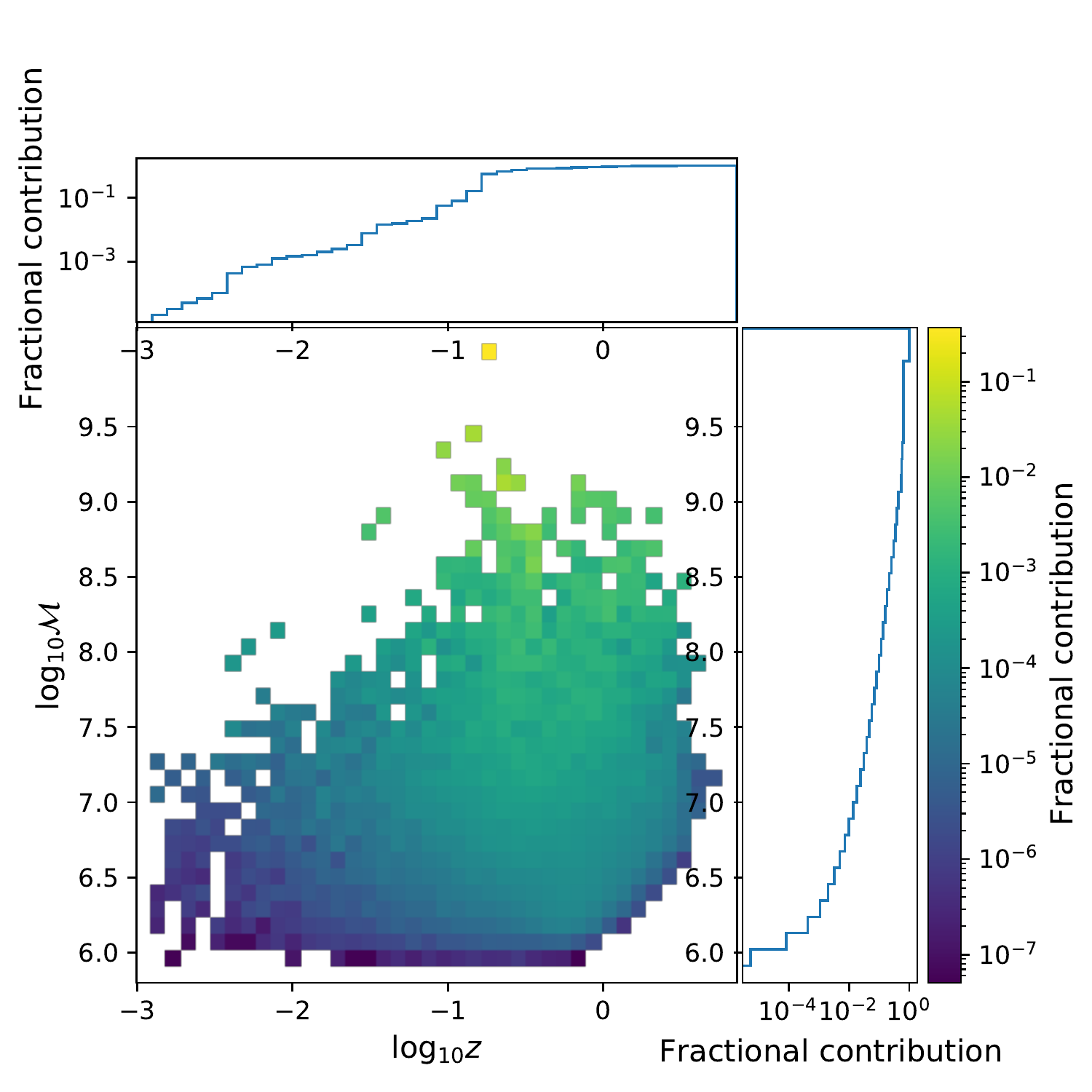}
\caption{GW, stellar scattering (SS) and gas disk (GD) with coalescence delay at $f = 1\,{\rm yr}^{-1}$}
\label{fig:contrib_hard_1yr}
\end{subfigure}

\caption{Contribution to $\Ayr$ for binaries binned by chirp mass and redshift assuming GW-driven binaries (top) and assuming binaries driven by stellar scattering and gas disk interactions also (bottom). White areas represent bins with no data. The colour scale represents the ratio of $h_{\rm c, bin}^{2}$ for binaries in each bin to the total $h_{\rm c}^{2}$. Marginalised cumulative distributions are given above and to the right of each subfigure and show the contribution to $h_{\rm c}^{2}$ for each row and column of bins. Note that the redshift axes of the top and bottom panel cover a different range.}
\label{fig:contribs_1yr}
\end{figure}

The contributions of SMBHBs to the SGWB are not equal across $\mathcal{M}$ and $z$. Figure~\ref{fig:contribs_1yr} shows the relative contributions to $\Ayr$ from mergers binned by $\mathcal{M}$ and $z$. For GW-driven mergers (Fig.~\ref{fig:contrib_gw_1yr}), the colour of each bin is determined by applying equation~\eqref{eqn:summation_hc} to the set of mergers in each bin and normalising by the total characteristic strain at $f = 1 \, {\rm yr}^{-1}$, $\Ayr$. In the case of mergers which include stellar scattering, gas disk interactions and coalescence delays (Fig.~\ref{fig:contrib_hard_1yr}), each SMBHB is evolved to the reference frequency of $1 \, {\rm yr}^{-1}$ whereupon the updated $\mathcal{M}$ and $z$, among other quantities, are used to calculate the contribution via equation~\eqref{eqn:summation_hardening}; the sums of the contributions in each bin are then normalised by $\Ayr$. Logarithmic bins are chosen to most evenly distribute the data and avoid an accumulation of mergers in low redshift bins, $z < 1$.

GW-driven merger contributions are shown in Fig.~\ref{fig:contrib_gw_1yr}. We find that $50\%$ of $\Ayr$ is produced by SMBHBs with $\mathcal{M} > 1.4 \times 10^{9} \, M_{\odot}$; there are four SMBHBs in this region: equivalent to $\sim 0.01\%$ of the total number of mergers. In addition, $35\%$ of $\Ayr$ can be attributed to just one merger. SMBHBs with $\mathcal{M} > 5 \times 10^{7} \, M_{\odot}$ are responsible for $90\%$ of the signal. These statistics suggest that low $\mathcal{M}$ binaries only very weakly contribute to $\Ayr$ with higher $\mathcal{M}$ binaries providing the majority of the signal. This is consistent with \citet{CornishSesana:2013} and \citet{WyitheLoeb:2003} who also find that nearby, massive binaries dominate the SGWB spectrum. These dominant SMBHBs are rare objects which illustrate the importance of a large-volume simulation where, by virtue of there being more mergers, statistically rare, high $\mathcal{M}$ binaries are more likely be found.

In terms of $z$, $50\%$ of $\Ayr$ is produced at $z < 0.2$ by $9\%$ of the SMBHB population. It is important to note that $z < 0.2$ includes several of the most massive SMBHBs. Using these observations, the sources can broadly be separated into a small set of massive, nearby SMBHBs and a much larger set of smaller and/or further away SMBHBs. The former includes the set of resolvable SMBHBs \citep{Boyle:2012} while the latter is a true stochastic background.

Similar statistics are obtained when considering the contributions obtained from equation~\eqref{eqn:summation_hardening}, shown in Fig.~\ref{fig:contrib_hard_1yr}. Again, $50\%$ of the SGWB is produced by those four massive mergers -- evidently environmental interactions and the coalescence delay do not remove or attenuate their contribution. As a fraction of the total $\Ayr$, the most massive merger now contributes $37\%$; this increase is likely a combination of a larger $\mathcal{M}$ due to accretion, and a reduction in the total $\Ayr$ due to SMBHBs which never evolve to the point of emitting GWs at $f = 1 \, {\rm yr}^{-1}$. There is a slight increase in the contribution of more massive SMBHBs with $90\%$ of the SGWB coming from binaries with $\mathcal{M} > 1 \times 10^{8} \, M_{\odot}$: a factor of $2$ increase in chirp mass from the GW-driven case. We may explain this in terms of the circumbinary gas disk primarily suppressing the contributions of low chirp mass binaries with little or no effect on high chirp mass binaries \citep{Kocsis:2011}.

Lower redshifts are seen in Fig.~\ref{fig:contrib_hard_1yr} compared to Fig.~\ref{fig:contrib_gw_1yr}, since SMBHBs are allowed to evolve closer to $z=0$. The plot is truncated at $z = 10^{-3}$ to aid in presentation -- this results in $43$ mergers being omitted in the figure; these are low-mass events which do not contribute significantly. There is a slight decrease in the redshift below which $50\%$ of $\Ayr$ is produced, to $z < 0.19$ ($z < 0.2$ in the GW-driven case). The obvious explanation is that the redshift of the entire coalescing population decreases when the binary formation-coalescence delay is introduced. We do note that $z = 0.19$ does fall roughly on the position of the most massive SMBHBs, and so the position of this $50\%$ marker is likely quite dependant on the redshift of those sources. Significantly, $50\%$ of the SMBHBs are at $z < 0.19$ compared to $9\%$ previously. $90\%$ of $\Ayr$ is produced at $z < 0.93$. The absence of data in the lower right (low chirp mass, high redshift) is a feature not seen in Fig.~ \ref{fig:contrib_gw_1yr}. The lack of GWs at $f = 1 \, {\rm yr}^{-1}$ for such SMBHBs hints at two possible scenarios: these SMBHBs are moving to lower redshift before emitting at $f = 1 \, {\rm yr}^{-1}$ due to the coalescence delay, or they are never reaching the binary separation corresponding to $f = 1 \, {\rm yr}^{-1}$. The latter in particular is expected for low chirp mass binaries since lower mass binaries tend to harden less quickly; this can be seen by inspection of equation~\eqref{eqn:t_df} where a lower satellite mass produces a longer infall time from dynamical friction. A combination of both effects is expected to produce the feature seen in Fig.~\ref{fig:contrib_hard_1yr}.

\subsection{Previous estimates}
\label{subsec:previous}
Here we compare our results to those obtained by the numerous other estimates over recent years. Since there are many such estimates, Table~\ref{tab:a_yr_v3} only includes a selection of them. First is the estimate by \citet{Kelley:2016} who use black hole data from the Illustris simulations to determine the characteristic strain in both the purely GW-driven case and the case where dynamical friction, stellar scattering and gas disk torques are simulated. The range given in the table corresponds to different hardening parameters. Work based on Enzo simulations by \citet{KulierCen:2015} produce an estimate for GW-driven binaries from a simulated black hole population; the range is due to statistical errors in this case, not due to producing estimates via different methods. Finally, \citet{WyitheLoeb:2003}, \citet{McWilliamsEtAl:2014} and \citet{Sesana:2013b} predict a value without utilising hydrodynamic simulations, instead using semi-analytic modelling.

In addition, Table~\ref{tab:a_yr_v3} contains the 95\% confidence upper limits for four PTA observatories. Unlike the other entries, these are not estimates of $\Ayr$ but are included to show consistency of our predictions, and others, with PTA observations. Our result for $\Ayr$ is below all PTA upper limits. Our estimates are also at least $2.3$ times lower than those based on Enzo simulations. It is difficult to isolate the exact cause of this since cosmological simulations are very complex and any number of their internal parameters could impact the SMBHB populations which develop. The results are also at least $6\%$ higher than the Illustris estimate, which again can be partially attributed to differences in the simulations themselves. Once the SGWB is detected, it could be used to judge the physical accuracy of simulations. The previous estimates of $\Ayr$ listed vary over an order of magnitude; our estimates fall inside this range, suggesting that our results agree with the existing literature.

As evidence of a SGWB detection increases, it will be valuable to compare estimates with these measurements. Compared to the recent NANOGrav result which has a median of $\Ayr = 1.92 \times 10^{-15}$ \citep{Nanograv_sep:2020}, our estimates are $2.8$ times lower in the GW-driven case and $3.0$ times lower in the GW and hardening model. More recently still, the PPTA \citep{PPTA:2021} makes an estimate with median $\Ayr = 2.2 \times 10^{-15}$. This is a factor of $3.2$ larger than our GW-driven model estimate and a factor of $3.4$ larger than our hardening model estimate. Lastly, the EPTA signal \citep{EPTA_2021} has a median of $\Ayr = 2.95 \times 10^{-15}$, $4.3$ and $4.6$ times larger than our GW-driven and hardening model estimates, respectively. If the signals prove to be the SGWB, these observations will help to inform future simulations and provide information about the population of SMBHBs \citep{Middleton:2021, Casey-Clyde:2021}.

\begin{table}
    \centering
    \begin{tabular}{|c|c|c|}
        \hline
         Notes & Source &  $A_{yr^{-1}}$ \\
         \hline
         Illustris & \citet{Kelley:2016} & $3-6 \times 10^{-16}$ \\
         Enzo & \citet{KulierCen:2015} & $2.0 \times 10^{-15}$ \\
         Semi-analytic & \citet{WyitheLoeb:2003} & $ 4.6 \times 10^{-15}$ \\
         Semi-analytic & \citet{McWilliamsEtAl:2014} & $4.1 \times 10^{-15}$ \\ 
         Semi-analytic & \citet{Sesana:2013b} & $3.5 - 15 \times 10^{-16}$ \\
         \hline
        NANOGrav & \citet{NANOGrav:2018} & $< 1.45 \times 10^{-15}$ \\
		PPTA & \citet{ShannonEtAlPPTA:2015} & $< 1.0 \times 10^{-15}$ \\
		EPTA & \citet{EPTASGWB:2015} & $< 3.0 \times 10^{-15}$ \\
		IPTA & \citet{IPTA:2016} & $< 1.7 \times 10^{-15}$ \\
		\hline
        GW only & \multirow{2}{*}{This work} & $6.9 \times 10^{-16}$ \\
		GW + SS + GD & & $6.4 \times 10^{-16}$ \\
		\hline
    \end{tabular}
    \caption{Top: estimates of the SGWB in the literature; middle: 95\% confidence upper limits as defined by each of the PTA groups; bottom: estimates as per this work.}
    \label{tab:a_yr_v3}
\end{table}

There is a difference of $\sim 8\%$ between the two estimates of $\Ayr$ presented in this work; that is, between the GW-driven estimate and the estimate where hardening mechanisms are taken into account and not all binaries coalesce. This percentage gap, which we will denote by $\Delta \Ayr$ for convenience, is much larger in other studies of a similar nature \citep{Kelley:2016, Siwek:2020}. \citet{Kelley:2016} estimate that these quantities differ by $20 - 130\%$ depending on the choice of hardening parameters (loss-cone refill rate, SMBH accretion rate, viscous disk parameters, etc.)

A similar comparison is provided by the estimates of \citet{Siwek:2020}, but in this case it is only the accretion model which is varied. These results still show significant variations of $24 - 300\%$ between non-accreting and accreting models. Using MB-II data we simulate a non-accreting model by manually setting $\dot{M}=0$. The analogous gap between $\Ayr$ for non-accreting and accreting models we find is $5\%$. Clearly there is some element of the analysis or data in our work which diminishes the gap between the two estimates compared to these other works.

At the reference frequency of $f = 1 \, {\rm yr}^{-1}$, the characteristic strain spectrum is very close in shape to the $f^{-2/3}$ power law predicted by binaries evolving by GW-emission only. Binaries which emit at this frequency will coalesce quickly due to GW emission, meaning that comparing the coalescing fraction for each binary evolution model may shed light on why $\Delta \Ayr$ differs. In our models, we find approximately $65\%$ of SMBHBs coalesce, whereas \citet{Kelley:2016} find a wide range of coalescence fractions from $15\%$ to $90\%$ depending on the choice of binary hardening parameters. These fractions are contrasted to the $100\%$ coalescence fraction of the corresponding GW-only estimates. \cite{Siwek:2020} also find variability in the coalescence fraction, ranging from $46.5\%$ to $70.8\%$ depending on how mass is accreted onto component SMBHs in the binary.

The difference between coalescing fractions found in these two works \citep{Kelley:2016, Siwek:2020} are of the same magnitude as those found here, tens of percent, which naively would suggest that $\Delta \Ayr$ should be similar in magnitude in all three cases. That there is large differences in $\Delta \Ayr$ between our results ($\sim 8\%$) and \citet{Kelley:2016} and \citet{Siwek:2020} ($20 - 300 \%$) indicates that the coalescence fraction alone is too coarse a measure and one must examine, not just the number, but also the types of binaries which are (or are not) coalescing to explain why $\Delta \Ayr$ varies between analyses.

Comparing Figures \ref{fig:contrib_gw_1yr} and \ref{fig:contrib_hard_1yr} show that the contributions of the most massive binaries are not strongly impacted by including hardening mechanisms in the calculation. With, for example, four binaries producing $50\%$ of $\Ayr$, these massive systems dominate the predicted signal; it is not surprising then that their lack of sizable change in mass or redshift between GW-only and hardening models due to accretion and time delay before coalescence corresponds to a small overall change in $\Ayr$. The SMBHB populations of \citet{Kelley:2016} and \citet{Siwek:2020} are smaller ($9270$ merger events) than that of MB-II ($43\,211$ merger events) and hence, rare, massive binaries such as the aforementioned four from MB-II are less likely in their merging populations. This may contribute to the larger differences between estimates in these other works, as SMBHBs with greater sensitivity to hardening mechanisms are contributing a larger fraction to the SGWB.

\section{Conclusions}
\label{sec:conclusion}
Data from the MB-II simulation are used to estimate the SGWB background from SMBHBs. MB-II is an excellent candidate for this due to being high-resolution, high-volume and utilising both N-body and SPH techniques. It also benefits from a set of sub-resolution physics models which differentiate it from existing simulations. Our prediction of the SGWB is computed in two ways, first by assuming SMBHB inspiral is driven purely by GW emission, and then again by including the effects of scattering with stars in the binary loss cone, torques with gas disks in gas-rich galaxy mergers, and delays between binary formation and coalescence. We find that these methods produce different predictions of the SGWB spectrum in the PTA band.

The estimate assuming a binary merger driven only by GW emission produces a SGWB signal for which $h_{\rm c}(f) \propto f^{-2/3}$ where $\Ayr = 6.9 \times 10^{-16}$. Including hardening mechanisms has a small but not insignificant effect, with our estimate yielding $\Ayr = 6.4 \times 10^{-16}$. These additional effects also result in a suppression of the SGWB with decreasing frequency, with a factor of $1.7$ reduction by $f = 10^{-9}\, {\rm Hz}$. This suppression may impact prospects of a PTA detection~\citep[see also][]{ShannonEtAlPPTA:2015}.

In Section~\ref{sec:SGWBcontributions} we find a strong dependence on chirp mass to the amount which a binary contributes to the total SGWB. It is noted that SMBHBs which are either nearby with $z < 0.2$, or with $\mathcal{M} > 1.4 \times 10^{9} M_{\odot}$, contribute a disproportionate amount, at least half, despite being less than half of the population. This is consistent with other works; for example, \citet{WyitheLoeb:2003}, \citet{CornishSesana:2013} and \citet{RaviEtAl:2015}. It is important to note that these very high chirp mass SMBHBs are statistically rare and as such, the population produced is subject to greater variability between hypothetical re-runs of the simulation code. This variability manifests as an inherent uncertainty of $0.5 - 0.8 \, {\rm dex}$ on the calculated characteristic strain.
Since the comoving volume of MB-II is significantly lower than the comoving volume of real space which would contain SMBHBs contributing to the SGWB, the expected value of $\Ayr$ will increase with increased simulation volume.

In this work we have assumed that SMBHBs have zero eccentricity. In reality this is unlikely as the binary will either begin with some eccentricity, or develop some as it interacts with nearby stars and gas. Accounting for eccentricity causes further suppression of $h_{\rm c}(f)$ at lower frequencies, where the exact turnover is determined by the amount of eccentricity \citep{Chen:2017}. Eccentricity data is not available for MB-II, however a more detailed treatment following \citep{Kelley:2017} may be possible. Further potential for uncertainties arise from the models which are used to describe binary evolution. The effects of three body interactions and diffusion of stars back into the loss cone of a SMBHB are represented by a very simplistic equation. Similarly, the circumbinary gas disk model we choose is based on an idealised gas disk model from \citet{Haiman:2009}. Additionally, we do not consider the different regimes in which gas disk interactions differ as in \citet{Kocsis:2011}. We do not include a $da/dt_{\rm r}$ term such as equations~\eqref{eqn:scattering} or~\eqref{eqn:circumbinary_disk} to account for the direct effect of dynamical friction in attenuating the SGWB spectrum. Instead, dynamical friction only influences the coalescence delay. The correction for treating dynamical friction in full rigour (i.e. with a corresponding time evolution equation) is expected to be very small.

We have contributed an estimate of the SGWB from MB-II to a growing collection in the literature. These predictions are useful when speculating the time until detection and comparisons with current upper limits provide a test for the underlying physical assumptions used in these estimates. Our results are consistent with PTA upper limits and are comparable to other predictions. Finding, or confirming, a SGWB will be a significant observational achievement in coming years, proving that SMBHBs form and merge and revealing the extent to which any hardening mechanisms come into play.

\section*{Data availability}
The data underlying this article were provided by the MB-II team. Data will be shared on request to the corresponding author with permission of MB-II team.

\section*{Acknowledgements}
% adapted from text at https://supercomputing.swin.edu.au/policies/
The authors are grateful to Ryan Shannon and the PPTA for use of the PPTA upper limit data for comparison. The authors also thank Siyuan Chen for valuable feedback.
This work used computational resources of the OzSTAR national facility at Swinburne University of Technology. 
The OzSTAR program receives funding in part from the Astronomy National Collaborative Research Infrastructure Strategy (NCRIS) allocation provided by the Australian Government.
% general OzGrav acknowledgement
This research is supported by the Australian Research Council Centre of Excellence for Gravitational Wave Discovery (OzGrav) (project number CE170100004).
% MB-II acknowledgement
The authors acknowledge the Texas Advanced Computing Center (TACC) at The
University of Texas at Austin for providing HPC resources that have
contributed to the research results reported within this paper. 
TDM acknowledges funding from NSF ACI-1614853,  NSF AST-1616168 and  
NASA ATP 80NSSC18K101, and NASA ATP NNX17AK56G.

%%%%%%%%%%%%%%%%%%%%%%%%%%%%%%%%%%%%%%%%%%%%%%%%%%



\bibliographystyle{mnras}
\bibliography{bibtexReferences}

\begin{thebibliography}{}
\makeatletter
\relax
\def\mn@urlcharsother{\let\do\@makeother \do\$\do\&\do\#\do\^\do\_\do\%\do\~}
\def\mn@doi{\begingroup\mn@urlcharsother \@ifnextchar [ {\mn@doi@}
  {\mn@doi@[]}}
\def\mn@doi@[#1]#2{\def\@tempa{#1}\ifx\@tempa\@empty \href
  {http://dx.doi.org/#2} {doi:#2}\else \href {http://dx.doi.org/#2} {#1}\fi
  \endgroup}
\def\mn@eprint#1#2{\mn@eprint@#1:#2::\@nil}
\def\mn@eprint@arXiv#1{\href {http://arxiv.org/abs/#1} {{\tt arXiv:#1}}}
\def\mn@eprint@dblp#1{\href {http://dblp.uni-trier.de/rec/bibtex/#1.xml}
  {dblp:#1}}
\def\mn@eprint@#1:#2:#3:#4\@nil{\def\@tempa {#1}\def\@tempb {#2}\def\@tempc
  {#3}\ifx \@tempc \@empty \let \@tempc \@tempb \let \@tempb \@tempa \fi \ifx
  \@tempb \@empty \def\@tempb {arXiv}\fi \@ifundefined
  {mn@eprint@\@tempb}{\@tempb:\@tempc}{\expandafter \expandafter \csname
  mn@eprint@\@tempb\endcsname \expandafter{\@tempc}}}

\bibitem[\protect\citeauthoryear{{Alam} et~al.,}{{Alam}
  et~al.}{2020}]{AlamEtAlNANOGrav:2020}
{Alam} M.~F.,  et~al., 2020, arXiv e-prints, \href
  {https://ui.adsabs.harvard.edu/abs/2020arXiv200506490A} {p. arXiv:2005.06490}

\bibitem[\protect\citeauthoryear{Antoniadis et~al.,}{Antoniadis
  et~al.}{2022}]{IPTA_2022}
Antoniadis J.,  et~al., 2022, \mn@doi [Monthly Notices of the Royal
  Astronomical Society] {10.1093/mnras/stab3418}

\bibitem[\protect\citeauthoryear{Armitage \& Natarajan}{Armitage \&
  Natarajan}{2002}]{Armitage:2002}
Armitage P.~J.,  Natarajan P.,  2002, \mn@doi [The Astrophysical Journal]
  {10.1086/339770}, 567, L9–L12

\bibitem[\protect\citeauthoryear{{Arzoumanian} et~al.,}{{Arzoumanian}
  et~al.}{2018}]{NANOGrav:2018}
{Arzoumanian} Z.,  et~al., 2018, \mn@doi [\apj] {10.3847/1538-4357/aabd3b},
  \href {https://ui.adsabs.harvard.edu/abs/2018ApJ...859...47A} {859, 47}

\bibitem[\protect\citeauthoryear{{Arzoumanian} et~al.,}{{Arzoumanian}
  et~al.}{2020}]{Nanograv_sep:2020}
{Arzoumanian} Z.,  et~al., 2020, arXiv e-prints, \href
  {https://ui.adsabs.harvard.edu/abs/2020arXiv200904496A} {p. arXiv:2009.04496}

\bibitem[\protect\citeauthoryear{Begelman, Blandford  \& Rees}{Begelman
  et~al.}{1980}]{Begelman:1980}
Begelman M.~C.,  Blandford R.~D.,   Rees M.~J.,  1980, Nature, 287, 307

\bibitem[\protect\citeauthoryear{{Bhowmick}, {DiMatteo}, {Eftekharzadeh}  \&
  {Myers}}{{Bhowmick} et~al.}{2019}]{Bhowmick:2019}
{Bhowmick} A.~K.,  {DiMatteo} T.,  {Eftekharzadeh} S.,   {Myers} A.~D.,  2019,
  \mn@doi [\mnras] {10.1093/mnras/stz519}, \href
  {https://ui.adsabs.harvard.edu/abs/2019MNRAS.485.2026B} {485, 2026}

\bibitem[\protect\citeauthoryear{{Binney} \& {Tremaine}}{{Binney} \&
  {Tremaine}}{2008}]{GalacticDynamicsBinney:2008}
{Binney} J.,  {Tremaine} S.,  2008, {Galactic Dynamics: Second Edition}.
Princeton University Press

\bibitem[\protect\citeauthoryear{Boylan-Kolchin, Ma  \&
  Quataert}{Boylan-Kolchin et~al.}{2007}]{Boylan:2007}
Boylan-Kolchin M.,  Ma C.-P.,   Quataert E.,  2007, \mn@doi [Monthly Notices of
  the Royal Astronomical Society] {10.1111/j.1365-2966.2007.12530.x}, 383, 93

\bibitem[\protect\citeauthoryear{Boyle \& Pen}{Boyle \& Pen}{2012}]{Boyle:2012}
Boyle L.,  Pen U.-L.,  2012, \mn@doi [Physical Review D]
  {10.1103/physrevd.86.124028}, 86

\bibitem[\protect\citeauthoryear{{Bryan} et~al.,}{{Bryan}
  et~al.}{2014}]{Enzo:2014}
{Bryan} G.~L.,  et~al., 2014, \mn@doi [\apjs] {10.1088/0067-0049/211/2/19},
  \href {https://ui.adsabs.harvard.edu/abs/2014ApJS..211...19B} {211, 19}

\bibitem[\protect\citeauthoryear{{Casey-Clyde}, {Mingarelli}, {Greene},
  {Pardo}, {Na{\~n}ez}  \& {Goulding}}{{Casey-Clyde}
  et~al.}{2021}]{Casey-Clyde:2021}
{Casey-Clyde} J.~A.,  {Mingarelli} C. M.~F.,  {Greene} J.~E.,  {Pardo} K.,
  {Na{\~n}ez} M.,   {Goulding} A.~D.,  2021, arXiv e-prints, \href
  {https://ui.adsabs.harvard.edu/abs/2021arXiv210711390C} {p. arXiv:2107.11390}

\bibitem[\protect\citeauthoryear{{Chen}, {Middleton}, {Sesana}, {Del Pozzo}  \&
  {Vecchio}}{{Chen} et~al.}{2017a}]{ChenEtAlInterp:2017}
{Chen} S.,  {Middleton} H.,  {Sesana} A.,  {Del Pozzo} W.,   {Vecchio} A.,
  2017a, \mn@doi [\mnras] {10.1093/mnras/stx475}, \href
  {https://ui.adsabs.harvard.edu/abs/2017MNRAS.468..404C} {468, 404}

\bibitem[\protect\citeauthoryear{Chen, Sesana  \& Del~Pozzo}{Chen
  et~al.}{2017b}]{Chen:2017}
Chen S.,  Sesana A.,   Del~Pozzo W.,  2017b, \mn@doi [Monthly Notices of the
  Royal Astronomical Society] {10.1093/mnras/stx1093}, 470, 1738–1749

\bibitem[\protect\citeauthoryear{{Chen}, {Sesana}  \& {Conselice}}{{Chen}
  et~al.}{2019}]{Chen:2019}
{Chen} S.,  {Sesana} A.,   {Conselice} C.~J.,  2019, \mn@doi [\mnras]
  {10.1093/mnras/stz1722}, \href
  {https://ui.adsabs.harvard.edu/abs/2019MNRAS.488..401C} {488, 401}

\bibitem[\protect\citeauthoryear{{Chen} et~al.,}{{Chen}
  et~al.}{2021}]{EPTA_2021}
{Chen} S.,  et~al., 2021, \mn@doi [\mnras] {10.1093/mnras/stab2833}, \href
  {https://ui.adsabs.harvard.edu/abs/2021MNRAS.508.4970C} {508, 4970}

\bibitem[\protect\citeauthoryear{{Cornish} \& {Sesana}}{{Cornish} \&
  {Sesana}}{2013}]{CornishSesana:2013}
{Cornish} N.~J.,  {Sesana} A.,  2013, \mn@doi [Classical and Quantum Gravity]
  {10.1088/0264-9381/30/22/224005}, \href
  {https://ui.adsabs.harvard.edu/abs/2013CQGra..30v4005C} {30, 224005}

\bibitem[\protect\citeauthoryear{{Dabringhausen}, {Hilker}  \&
  {Kroupa}}{{Dabringhausen} et~al.}{2008}]{Dabringhausen:2008}
{Dabringhausen} J.,  {Hilker} M.,   {Kroupa} P.,  2008, \mn@doi [\mnras]
  {10.1111/j.1365-2966.2008.13065.x}, \href
  {https://ui.adsabs.harvard.edu/abs/2008MNRAS.386..864D} {386, 864}

\bibitem[\protect\citeauthoryear{DeGraf, Di~Matteo, Treu, Feng, Woo  \&
  Park}{DeGraf et~al.}{2015}]{DeGraf:2015}
DeGraf C.,  Di~Matteo T.,  Treu T.,  Feng Y.,  Woo J.-H.,   Park D.,  2015,
  \mn@doi [Monthly Notices of the Royal Astronomical Society]
  {10.1093/mnras/stv2002}, 454, 913

\bibitem[\protect\citeauthoryear{{Dehnen}}{{Dehnen}}{1993}]{Dehnen:1993}
{Dehnen} W.,  1993, \mn@doi [\mnras] {10.1093/mnras/265.1.250}, \href
  {https://ui.adsabs.harvard.edu/abs/1993MNRAS.265..250D} {265, 250}

\bibitem[\protect\citeauthoryear{{Desvignes} et~al.,}{{Desvignes}
  et~al.}{2016}]{DesvignesEtAlEPTA:2016}
{Desvignes} G.,  et~al., 2016, \mn@doi [\mnras] {10.1093/mnras/stw483}, \href
  {https://ui.adsabs.harvard.edu/abs/2016MNRAS.458.3341D} {458, 3341}

\bibitem[\protect\citeauthoryear{{Dosopoulou} \& {Antonini}}{{Dosopoulou} \&
  {Antonini}}{2017}]{Dosopoulou:2017}
{Dosopoulou} F.,  {Antonini} F.,  2017, \mn@doi [\apj]
  {10.3847/1538-4357/aa6b58}, \href
  {https://ui.adsabs.harvard.edu/abs/2017ApJ...840...31D} {840, 31}

\bibitem[\protect\citeauthoryear{Dotti, Sesana  \& Decarli}{Dotti
  et~al.}{2012}]{Dotti:2012}
Dotti M.,  Sesana A.,   Decarli R.,  2012, \mn@doi [Advances in Astronomy]
  {10.1155/2012/940568}, 2012, 1–14

\bibitem[\protect\citeauthoryear{{Foster} \& {Backer}}{{Foster} \&
  {Backer}}{1990}]{FosterBacker:1990}
{Foster} R.~S.,  {Backer} D.~C.,  1990, \mn@doi [\apj] {10.1086/169195}, \href
  {https://ui.adsabs.harvard.edu/abs/1990ApJ...361..300F} {361, 300}

\bibitem[\protect\citeauthoryear{{Goncharov} et~al.,}{{Goncharov}
  et~al.}{2021}]{PPTA:2021}
{Goncharov} B.,  et~al., 2021, arXiv e-prints, \href
  {https://ui.adsabs.harvard.edu/abs/2021arXiv210712112G} {p. arXiv:2107.12112}

\bibitem[\protect\citeauthoryear{Haiman, Kocsis  \& Menou}{Haiman
  et~al.}{2009}]{Haiman:2009}
Haiman Z.,  Kocsis B.,   Menou K.,  2009, \mn@doi [The Astrophysical Journal]
  {10.1088/0004-637x/700/2/1952}, 700, 1952–1969

\bibitem[\protect\citeauthoryear{Hilborn}{Hilborn}{2018}]{Hilborn:2018}
Hilborn R.~C.,  2018, \mn@doi [American Journal of Physics]
  {10.1119/1.5020984}, 86, 186–197

\bibitem[\protect\citeauthoryear{{Ivanov}, {Papaloizou}  \&
  {Polnarev}}{{Ivanov} et~al.}{1999}]{Ivanov:1999}
{Ivanov} P.~B.,  {Papaloizou} J.~C.~B.,   {Polnarev} A.~G.,  1999, \mn@doi
  [\mnras] {10.1046/j.1365-8711.1999.02623.x}, \href
  {https://ui.adsabs.harvard.edu/abs/1999MNRAS.307...79I} {307, 79}

\bibitem[\protect\citeauthoryear{{Joshi} et~al.,}{{Joshi}
  et~al.}{2018}]{JoshiInPTA:2018}
{Joshi} B.~C.,  et~al., 2018, \mn@doi [Journal of Astrophysics and Astronomy]
  {10.1007/s12036-018-9549-y}, \href
  {https://ui.adsabs.harvard.edu/abs/2018JApA...39...51J} {39, 51}

\bibitem[\protect\citeauthoryear{Kelley, Blecha  \& Hernquist}{Kelley
  et~al.}{2016}]{Kelley:2016}
Kelley L.~Z.,  Blecha L.,   Hernquist L.,  2016, \mn@doi [Monthly Notices of
  the Royal Astronomical Society] {10.1093/mnras/stw2452}, 464, 3131

\bibitem[\protect\citeauthoryear{Kelley, Blecha, Hernquist, Sesana  \&
  Taylor}{Kelley et~al.}{2017}]{Kelley:2017}
Kelley L.~Z.,  Blecha L.,  Hernquist L.,  Sesana A.,   Taylor S.~R.,  2017,
  \mn@doi [Monthly Notices of the Royal Astronomical Society]
  {10.1093/mnras/stx1638}, 471, 4508–4526

\bibitem[\protect\citeauthoryear{{Khandai}, {Di Matteo}, {Croft}, {Wilkins},
  {Feng}, {Tucker}, {DeGraf}  \& {Liu}}{{Khandai}
  et~al.}{2015}]{MBIIKhandaiEtAl:2015}
{Khandai} N.,  {Di Matteo} T.,  {Croft} R.,  {Wilkins} S.,  {Feng} Y.,
  {Tucker} E.,  {DeGraf} C.,   {Liu} M.-S.,  2015, \mn@doi [\mnras]
  {10.1093/mnras/stv627}, \href
  {https://ui.adsabs.harvard.edu/abs/2015MNRAS.450.1349K} {450, 1349}

\bibitem[\protect\citeauthoryear{Kocsis \& Sesana}{Kocsis \&
  Sesana}{2011}]{Kocsis:2011}
Kocsis B.,  Sesana A.,  2011, \mn@doi [Monthly Notices of the Royal
  Astronomical Society] {10.1111/j.1365-2966.2010.17782.x}, 411, 1467–1479

\bibitem[\protect\citeauthoryear{Komatsu et~al.,}{Komatsu et~al.}{2011}]{WMAP7}
Komatsu E.,  et~al., 2011, \mn@doi [The Astrophysical Journal Supplement
  Series] {10.1088/0067-0049/192/2/18}, 192, 18

\bibitem[\protect\citeauthoryear{{Kormendy} \& {Ho}}{{Kormendy} \&
  {Ho}}{2013}]{KormendyHo:2013}
{Kormendy} J.,  {Ho} L.~C.,  2013, \mn@doi [\araa]
  {10.1146/annurev-astro-082708-101811}, \href
  {https://ui.adsabs.harvard.edu/abs/2013ARA&A..51..511K} {51, 511}

\bibitem[\protect\citeauthoryear{{Kulier}, {Ostriker}, {Natarajan}, {Lackner}
  \& {Cen}}{{Kulier} et~al.}{2015}]{KulierCen:2015}
{Kulier} A.,  {Ostriker} J.~P.,  {Natarajan} P.,  {Lackner} C.~N.,   {Cen} R.,
  2015, \mn@doi [\apj] {10.1088/0004-637X/799/2/178}, \href
  {https://ui.adsabs.harvard.edu/abs/2015ApJ...799..178K} {799, 178}

\bibitem[\protect\citeauthoryear{{Lentati} et~al.,}{{Lentati}
  et~al.}{2015}]{EPTASGWB:2015}
{Lentati} L.,  et~al., 2015, \mn@doi [\mnras] {10.1093/mnras/stv1538}, \href
  {https://ui.adsabs.harvard.edu/abs/2015MNRAS.453.2576L} {453, 2576}

\bibitem[\protect\citeauthoryear{L{\'o}pez, Degraf, DiMatteo, Fu, Fink  \&
  Gibson}{L{\'o}pez et~al.}{2011}]{Lopez:2011}
L{\'o}pez J.,  Degraf C.,  DiMatteo T.,  Fu B.,  Fink E.,   Gibson G.,  2011,
  in Bayard~Cushing J.,  French J.,   Bowers S.,  eds, Scientific and
  Statistical Database Management. Springer Berlin Heidelberg, Berlin,
  Heidelberg, pp 546--554

\bibitem[\protect\citeauthoryear{{Makino} \& {Funato}}{{Makino} \&
  {Funato}}{2004}]{Makino:2004}
{Makino} J.,  {Funato} Y.,  2004, \mn@doi [\apj] {10.1086/380917}, \href
  {https://ui.adsabs.harvard.edu/abs/2004ApJ...602...93M} {602, 93}

\bibitem[\protect\citeauthoryear{{McConnell} \& {Ma}}{{McConnell} \&
  {Ma}}{2013}]{McConnell_Ma:2013}
{McConnell} N.~J.,  {Ma} C.-P.,  2013, \mn@doi [\apj]
  {10.1088/0004-637X/764/2/184}, \href
  {https://ui.adsabs.harvard.edu/abs/2013ApJ...764..184M} {764, 184}

\bibitem[\protect\citeauthoryear{{McWilliams}, {Ostriker}  \&
  {Pretorius}}{{McWilliams} et~al.}{2014}]{McWilliamsEtAl:2014}
{McWilliams} S.~T.,  {Ostriker} J.~P.,   {Pretorius} F.,  2014, \mn@doi [\apj]
  {10.1088/0004-637X/789/2/156}, \href
  {https://ui.adsabs.harvard.edu/abs/2014ApJ...789..156M} {789, 156}

\bibitem[\protect\citeauthoryear{{Middleton}, {Del Pozzo}, {Farr}, {Sesana}  \&
  {Vecchio}}{{Middleton} et~al.}{2016}]{Middleton:2016}
{Middleton} H.,  {Del Pozzo} W.,  {Farr} W.~M.,  {Sesana} A.,   {Vecchio} A.,
  2016, \mn@doi [\mnras] {10.1093/mnrasl/slv150}, \href
  {https://ui.adsabs.harvard.edu/abs/2016MNRAS.455L..72M} {455, L72}

\bibitem[\protect\citeauthoryear{{Middleton}, {Chen}, {Del Pozzo}, {Sesana}  \&
  {Vecchio}}{{Middleton} et~al.}{2018}]{MiddletonEtAl:2018}
{Middleton} H.,  {Chen} S.,  {Del Pozzo} W.,  {Sesana} A.,   {Vecchio} A.,
  2018, \mn@doi [Nature Communications] {10.1038/s41467-018-02916-7}, \href
  {https://ui.adsabs.harvard.edu/abs/2018NatCo...9..573M} {9, 573}

\bibitem[\protect\citeauthoryear{{Middleton}, {Sesana}, {Chen}, {Vecchio}, {Del
  Pozzo}  \& {Rosado}}{{Middleton} et~al.}{2021}]{Middleton:2021}
{Middleton} H.,  {Sesana} A.,  {Chen} S.,  {Vecchio} A.,  {Del Pozzo} W.,
  {Rosado} P.~A.,  2021, \mn@doi [\mnras] {10.1093/mnrasl/slab008}, \href
  {https://ui.adsabs.harvard.edu/abs/2021MNRAS.502L..99M} {502, L99}

\bibitem[\protect\citeauthoryear{Milosavljevi\'{c}}{Milosavljevi\'{c}}{2003}]{Milosavljevic:2003}
Milosavljevi\'{c} M.,  2003, \mn@doi [AIP Conference Proceedings]
  {10.1063/1.1629432}

\bibitem[\protect\citeauthoryear{Nelson et~al.,}{Nelson
  et~al.}{2015}]{Illustris:2015}
Nelson D.,  et~al., 2015, \mn@doi [Astronomy and Computing]
  {10.1016/j.ascom.2015.09.003}, 13, 12–37

\bibitem[\protect\citeauthoryear{{Perera} et~al.,}{{Perera}
  et~al.}{2019}]{PereraEtAlIPTA:2019}
{Perera} B.~B.~P.,  et~al., 2019, \mn@doi [\mnras] {10.1093/mnras/stz2857},
  \href {https://ui.adsabs.harvard.edu/abs/2019MNRAS.490.4666P} {490, 4666}

\bibitem[\protect\citeauthoryear{{Phinney}}{{Phinney}}{2001}]{Phinney:2001}
{Phinney} E.~S.,  2001, arXiv e-prints, \href
  {https://ui.adsabs.harvard.edu/abs/2001astro.ph..8028P} {pp
  astro--ph/0108028}

\bibitem[\protect\citeauthoryear{{Ravi}, {Wyithe}, {Shannon}  \&
  {Hobbs}}{{Ravi} et~al.}{2015}]{RaviEtAl:2015}
{Ravi} V.,  {Wyithe} J.~S.~B.,  {Shannon} R.~M.,   {Hobbs} G.,  2015, \mn@doi
  [\mnras] {10.1093/mnras/stu2659}, \href
  {https://ui.adsabs.harvard.edu/abs/2015MNRAS.447.2772R} {447, 2772}

\bibitem[\protect\citeauthoryear{{Ryu}}{{Ryu}}{2018}]{Ryu:2018}
{Ryu} T.,  2018, PhD thesis, State University of New York at Stony Brook

\bibitem[\protect\citeauthoryear{{Schaye} et~al.,}{{Schaye}
  et~al.}{2015}]{SchayeCrain:2015}
{Schaye} J.,  et~al., 2015, \mn@doi [\mnras] {10.1093/mnras/stu2058}, \href
  {https://ui.adsabs.harvard.edu/abs/2015MNRAS.446..521S} {446, 521}

\bibitem[\protect\citeauthoryear{{Sesana}}{{Sesana}}{2013a}]{Sesana:2013a}
{Sesana} A.,  2013a, \mn@doi [Classical and Quantum Gravity]
  {10.1088/0264-9381/30/22/224014}, \href
  {https://ui.adsabs.harvard.edu/abs/2013CQGra..30v4014S} {30, 224014}

\bibitem[\protect\citeauthoryear{Sesana}{Sesana}{2013b}]{Sesana:2013b}
Sesana A.,  2013b, \mn@doi [Monthly Notices of the Royal Astronomical Society:
  Letters] {10.1093/mnrasl/slt034}, 433, L1–L5

\bibitem[\protect\citeauthoryear{Sesana \& Khan}{Sesana \&
  Khan}{2015}]{Sesana:2015}
Sesana A.,  Khan F.~M.,  2015, \mn@doi [Monthly Notices of the Royal
  Astronomical Society: Letters] {10.1093/mnrasl/slv131}, 454, L66

\bibitem[\protect\citeauthoryear{Sesana, Vecchio  \& Volonteri}{Sesana
  et~al.}{2009}]{Sesana:2009}
Sesana A.,  Vecchio A.,   Volonteri M.,  2009, \mn@doi [Monthly Notices of the
  Royal Astronomical Society] {10.1111/j.1365-2966.2009.14499.x}, 394,
  2255–2265

\bibitem[\protect\citeauthoryear{{Shannon} et~al.,}{{Shannon}
  et~al.}{2015}]{ShannonEtAlPPTA:2015}
{Shannon} R.~M.,  et~al., 2015, \mn@doi [Science] {10.1126/science.aab1910},
  \href {https://ui.adsabs.harvard.edu/abs/2015Sci...349.1522S} {349, 1522}

\bibitem[\protect\citeauthoryear{{Siwek}, {Kelley}  \& {Hernquist}}{{Siwek}
  et~al.}{2020}]{Siwek:2020}
{Siwek} M.~S.,  {Kelley} L.~Z.,   {Hernquist} L.,  2020, \mn@doi [\mnras]
  {10.1093/mnras/staa2361}, \href
  {https://ui.adsabs.harvard.edu/abs/2020MNRAS.498..537S} {498, 537}

\bibitem[\protect\citeauthoryear{{Springel}}{{Springel}}{2005}]{Springel:2005}
{Springel} V.,  2005, \mn@doi [\mnras] {10.1111/j.1365-2966.2005.09655.x},
  \href {https://ui.adsabs.harvard.edu/abs/2005MNRAS.364.1105S} {364, 1105}

\bibitem[\protect\citeauthoryear{Springel et~al.,}{Springel
  et~al.}{2005}]{Millenium:2005}
Springel V.,  et~al., 2005, \mn@doi [Nature] {10.1038/nature03597}, 435,
  629–636

\bibitem[\protect\citeauthoryear{{Stickley}}{{Stickley}}{2013}]{Stickley:2013}
{Stickley} N.~R.,  2013, PhD thesis, University of California, Riverside

\bibitem[\protect\citeauthoryear{Taylor, Vallisneri, Ellis, Mingarelli, Lazio
  \& van Haasteren}{Taylor et~al.}{2016}]{Taylor:2016}
Taylor S.~R.,  Vallisneri M.,  Ellis J.~A.,  Mingarelli C. M.~F.,  Lazio T.
  J.~W.,   van Haasteren R.,  2016, \mn@doi [The Astrophysical Journal]
  {10.3847/2041-8205/819/1/l6}, 819, L6

\bibitem[\protect\citeauthoryear{{Verbiest} et~al.,}{{Verbiest}
  et~al.}{2016}]{IPTA:2016}
{Verbiest} J.~P.~W.,  et~al., 2016, \mn@doi [\mnras] {10.1093/mnras/stw347},
  \href {https://ui.adsabs.harvard.edu/abs/2016MNRAS.458.1267V} {458, 1267}

\bibitem[\protect\citeauthoryear{{White} \& {Rees}}{{White} \&
  {Rees}}{1978}]{WhiteRees:1978}
{White} S.~D.~M.,  {Rees} M.~J.,  1978, \mn@doi [\mnras]
  {10.1093/mnras/183.3.341}, \href
  {https://ui.adsabs.harvard.edu/abs/1978MNRAS.183..341W} {183, 341}

\bibitem[\protect\citeauthoryear{{Wyithe} \& {Loeb}}{{Wyithe} \&
  {Loeb}}{2003}]{WyitheLoeb:2003}
{Wyithe} J. S.~B.,  {Loeb} A.,  2003, \mn@doi [\apj] {10.1086/375187}, \href
  {https://ui.adsabs.harvard.edu/abs/2003ApJ...590..691W} {590, 691}

\makeatother
\end{thebibliography}

%\begin{thebibliography}{99}
%\end{thebibliography}

%%%%%%%%%%%%%%%%%%%%%%%%%%%%%%%%%%%%%%%%%%%%%%%%%%

%%%%%%%%%%%%%%%%% APPENDICES %%%%%%%%%%%%%%%%%%%%%

%%%%%%%%%%%%%%%%%%%%%%%%%%%%%%%%%%%%%%%%%%%%%%%%%%

% Don't change these lines
\bsp	% typesetting comment
\label{lastpage}
\end{document}